\documentclass{stavanger}

\usepackage[utf8]{inputenc}
\usepackage{amssymb,amsmath,amsfonts,amsthm}
\usepackage{hyperref}
\usepackage{natbib}
\usepackage{graphicx}
\usepackage{tikz}
\usepackage{cleveref}
\usepackage{mathtools}
\usepackage{placeins}
\usepackage[export]{adjustbox} 

\startlocaldefs

\newcommand{\lt}{\tau_{\rm L}} 
\newcommand{\rt}{x_0}    
\newcommand{\f}{\phi}    

\endlocaldefs

\begin{document}


\begin{frontmatter}

\begin{fmbox}


\dochead{Research Article - Preprint}{FP}


\title{Stable solvers for real-time Complex Langevin }


\author[
   addressref={aff1},                   	  
   corref={aff1},                     		  
   email={daniel.alvestad@uis.no}   		  
]{\inits{DA}\fnm{Daniel} \snm{Alvestad}}
\author[
   addressref={aff1},                   	  
   email={rasmus.n.larsen@uis.no}   		  
]{\inits{RL}\fnm{Rasmus} \snm{Larsen}}
\author[
   addressref={aff1},                   	  
   email={alexander.rothkopf@uis.no}   		  
]{\inits{AR}\fnm{Alexander} \snm{Rothkopf}}


\address[id=aff1]{
  \orgname{Faculty of Science and Technology}, 	 
  \street{University of Stavanger},                     		 
  \postcode{4021}                               			 
  \city{Stavanger},                              				 
  \cny{Norway}                                   				 
}



\end{fmbox}


\begin{abstractbox}

\begin{abstract} 
This study explores the potential of modern implicit solvers for stochastic partial differential equations in the simulation of real-time complex Langevin dynamics. Not only do these methods offer asymptotic stability, rendering the issue of runaway solution moot, but they also allow us to simulate at comparatively large Langevin time steps, leading to lower computational cost. We compare different ways of regularizing the underlying path integral and estimate the errors introduced due to the finite Langevin time. Based on that insight, we implement benchmark (non-)thermal simulations of the quantum anharmonic oscillator on the canonical Schwinger-Keldysh contour of short real-time extent.
\end{abstract}



\end{abstractbox}


\end{frontmatter}


\section{Motivation}
\label{sec:introduction}
The sign problem (see ref.~\cite{Gattringer:2016kco} for a mini-review) remains one of the central open challenges in modern theoretical physics and hinders progress in various different subfields. It underlies the challenges encountered in the study of transport properties of the quark-gluon-plasma \cite{Meyer:2007ic, Amato:2013naa, Brandt:2015aqk, Ding:2016hua, Astrakhantsev:2017nrs, Astrakhantsev:2018oue, Astrakhantsev:2019zkr}, it is the central hurdle in the exploration of the QCD phase diagram at large Baryon density \cite{Bellwied:2015rza,Bazavov:2017dus,Borsanyi:2018grb,Scherzer:2020kiu, Attanasio:2020spv} and impacts the study of the thermodynamics of imbalanced Fermi gases \cite{Chevy:2010zz, Braun:2012ww, Gubbels:2013mda}, to name just three. The term sign problem refers to the fact that many strongly correlated quantum systems of phenomenological relevance can only be expressed through a path integral with complex valued Feynman weight. In turn, Monte-Carlo sampling methods, successful in case that the Feynman weight is purely real, become inapplicable and system-specific strategies must be developed. One of the most technically challenging sign problems occurs in case of quantum systems formulated in Minkowski spacetime, where the Feynman weight amounts to a pure phase.

The sign problem has been shown to be NP-hard \cite{Troyer:2004ge}, which implies that a one-size-fits-all approach is unlikely to exist. Nevertheless, many examples are known in which the sign problem has been successfully overcome or at least tamed. An active research community (for a recent review see ref.~\cite{Berger:2019odf}) is exploring multiple strategies. Among these are variants of reweighing, extrapolation from complex parameters and the reformulation of the system of interest in new degrees of freedom unaffected by a sign problem. 

The present study sets out to contribute to ongoing efforts to beat the sign problem by considering the complexification of system degrees of freedom. This research field has a long history, giving birth to two major promising currents: the Lefshets thimble approach \cite{Cristoforetti:2012su} and complex Langevin \cite{Seiler:2017wvd}. In the former, one identifies manifolds in the complex plane, the so-called thimbles, on which the imaginary part of the classical action remains constant and thus ordinary Monte-Carlo sampling may commence. A residual sign problem persists as one has to average over different thimbles. Together with the computationally demanding task of locating the thimbles, these challenges constitute two areas of active research interest. 

Complex Langevin on the other hand is based on the concept of stochastic quantization \cite{Parisi:1980ys,Damgaard:1987rr}. Quantum and statistical fluctuations of a system are represented by noise in an additional $(d+1)+1$ temporal dimension, which reproduces the correlation functions in $(d+1)$ dimensions. In practice one is required to evolve the field degrees of freedom by a stochastic partial differential equation (SDE) in an additional, so-called Langevin time, generating representations of the quantum system along the way. Expectation values of observables are estimated by taking the mean over these field configurations. Langevin stochastic quantization has proven successful in systems in which naive Monte-Carlo methods are also applicable \cite{Batrouni1985}. It has furthermore been shown to correctly simulate several systems with complex weights and thus complexified field degrees of freedom \cite{Seiler:2017wvd}. Even though straight forward in principle, it has been realized early on by the community that in its standard formulation, the complex Langevin approach suffers from three major shortcomings, which have to be addressed, before the method can serve as a reliable tool in the precision study of strongly correlated quantum systems.

The three main challenges affecting complex Langevin identified in the literature are its \textit{stability}, the \textit{ergodicity in the presence of non-holomorphic actions} and most crucially the \textit{convergence to incorrect results} (for recent insight see e.g. \cite{Scherzer:2018hid,Scherzer:2019lrh}. We believe that it is paramount to disentangle each of these issues, in order to be able to solve them one-by-one. Hence we focus in this study solely on the question of stability, returning to the remaining two in future work. (I.e. in order to remain in the parameter range where the complex Langevin method itself is known to converge to the correct results, we limit ourselves to a short real-time extent in this study.)

The question of stability in complex Langevin is intimately connected to the well-known phenomenon of \textit{runaway solutions}. In general, such divergent behavior can arise from two sources. Either the complex Langevin method itself does not converge to a finite result, or the numerical methods used to implement the discrete Langevin time evolution introduce artifacts, which in turn give rise to unphysical divergencies. In order to make progress on understanding the former, we must disentangle numerical artifacts from methods artifacts.

Observed early on \cite{Flower1986}, runaways are now commonly treated by deploying adaptive step-size prescriptions \cite{AartsJames2010} in the solution of the stochastic Langevin dynamics. One motivation for our work is the fact that even though adaptive step-size has proven to alleviate the problem of runaways in many systems in practice, it does not prevent their occurrence in principle. I.e. runaway solutions may appear even if adaptive step-size is deployed (see e.g. \cite{Seiler2013}).

In this paper, we approach the stability of complex Langevin from the point of view of the stiffness of the underlying SDEs. While no precise definition of stiffness exists, we take the pragmatic view that it refers to systems in which naive explicit time-stepping prescriptions fail to recover the correct solution. Surveying the landscape of complex Langevin implementations, we find that a majority of studies rely on the simple forward Euler discretization. Early on, improvements in the spirit of deterministic Runge-Kutta methods have been proposed \cite{Kronfeld:1992jf}, but to our knowledge only a single study \cite{Aarts2012} has embraced a higher-order method that takes into account the stochastic character of the Langevin evolution equation. 

Our study aims at bringing to the table some of the progress made in the solution of SDEs in other fields. In particular, we propose to deploy implicit solvers, which are designed with stiff problems in mind. Besides the simple  Euler-Maruyama scheme, which we use extensively in this paper, we will discuss what ingredients are needed in order to set up higher-order schemes for SDEs, compared to the case of purely deterministic equations. 

Once stable solvers are available, we can proceed to investigate the stability and accuracy of the complex Langevin method itself. Our goal lies in simulating real-time physics, which in the continuum requires a form of regularization. After exploring different ways how a regularization may be incorporated in the Langevin evolution, we implement high accuracy simulations of the (0+1) dimensional anharmonic oscillator in thermal equilibrium and as a genuine initial value problem from a Gaussian density matrix.

The paper is organized in the following way: we start in \cref{sec:langevin_equation} with an introduction of the equations underlying the complex Langevin approach and discuss some explicit and implicit SDE solvers for their solution. In the following, we prepare the grounds for numerical simulations by introducing and discretizing the anharmonic oscillator model in \cref{sec:model}. We discuss different ways to regularize its path integral \cref{sec:regularizing_CLE} and will learn how to describe the errors made by a finite step size in the Langevin evolution. Armed with this insight, we carry out benchmark complex Langevin simulations of the quantum anharmonic oscillator on the canonical Schwinger-Keldysh contour at short real-times both in thermal equilibrium and in a non-equilibrium setting in \cref{sec:numsim}. We close with a summary and outlook in \cref{sec:sumout}.
 

\section{Complex Langevin and SDE solvers}
\label{sec:langevin_equation}

The task at hand is to compute quantum statistical expectation values of an observable $O$. Conventionally such expectation values are formulated in terms of a Feynman path integral 
\begin{align}
    \langle O \rangle = \frac{1}{Z} \int {\cal D}\phi\;O[\phi] e^{iS[\phi]}, \quad S[\phi]=\int d^dx L[\phi].
\end{align}
where $Z=\int {\cal D}\phi \,{\rm exp}[iS[\phi]]$ denotes the partition function. 

In \textit{Stochastic Quantization}, we obtain the expectation values from the evolution of the system in an artificial Langevin time $\tau_{\rm L}$ (for an in-depth review of the approach, see Ref.\cite{Namiki:1992wf}). The Langevin-like evolution equation for the field $\phi(x,\tau_{\rm L})$ in its simplest form consists of a drift term, derived from its classical action $S[\phi]$, as well as a Gaussian noise term $\eta(x,\lt)$
\begin{equation}\label{eq:CLE}
\begin{aligned}
    & \frac{d\f}{d\lt} = i\frac{\delta S[\f]}{\delta \f(x)} + \eta(x,\lt) \quad \textrm{with}   \\
    &\langle \eta(x,\lt) \rangle = 0, \quad \langle \eta(x,\lt) \eta(x',\lt') \rangle = 2\delta(x-x')\delta(\lt-\lt').
\end{aligned}
\end{equation}
The quantity $x$ may e.g. refer to a four-vector $x=(x_0,\mathbf{x})$ with Minkowski (real-time) $x_0$ and the 3 spatial dimensions $\mathbf{x}$. It is important to keep in mind that the physical time ($x_0$) and the fictitious Langevin time $\lt$ are not related to each other. Note that the delta function in the correlator of the noise encompasses all dimensions of x. This prescription places an independent stochastic process at each space-time point. 

Due to the complex drift term, the field degrees complexify and we can rewrite the evolution equations instead in terms of the real and imaginary part of the field as
\begin{equation}
    \f(x,\lt) = \f_{R}(x,\lt) + i\f_I(x,\lt) ,
\end{equation}
such that \cref{eq:CLE} turns into two coupled but real-valued equations for the real- and imaginary part of the field degrees of freedom
\begin{equation}\label{eq:CLE_real}
\begin{aligned}
    & \frac{d\f_R}{d\lt} = \textrm{Re}\left[ \left. i\frac{\delta S[\f]}{\delta \f(x)}\right|_{\f = \f_R + i\f_I} \right] + \eta(x,\lt), \quad \frac{d\f_I}{d\lt} = \textrm{Im}\left[ \left. i\frac{\delta S[\f]}{\delta \f(x)}\right|_{\f = \f_R + i\f_I} \right]. \\
\end{aligned}
\end{equation}
We have here used the standard construction in which the noise term $\eta(c,\lt)$ is real. These are the stochastic partial differential evolution equations we will solve in the subsequent sections. The first central task is to find appropriate numerical solvers to accommodate these equations.

\subsection{Numerical schemes}\label{sec:schemes}

Stochastic partial differential equations (SDE) are a central modern tool in the modeling of various phenomena in science, technology and in particular finance. Most of the equations arising in these research fields do not lend themselves to an analytic treatment and thus require numerical solvers. To this end, the past two decades have seen vigorous research activity in the development of accurate and efficient algorithms. One major impulse towards these developments can be found in the by now classic book by Kl\"oden and Platen \cite{KloedenPlaten}, which not only contains a comprehensive survey of both explicit and implicit SDE solvers but also provides a pedagogic introduction into the underlying Ito and Stratonovic calculus. One central message of the book states that the series expansions, commonly used to set up deterministic discretization schemes, need to be amended by additional terms in the stochastic case due to the different scaling properties of stochastic variables. In particular, it is shown that deterministic algorithms may yield much lower convergence rates in an SDE setting than for the PDEs they were originally designed for. 

The goal of this section is to introduce some of these numerical schemes and to explicitly match the complex Langevin equations to the mathematical notation used in the literature, preparing us for a straightforward implementation through standard libraries, such as the SDE solver package found in the Julia language.

Let us formulate a general stochastic differential equation for $N$ stochastic variables $\f^i$ in Langevin time $\lt$, enumerated by the superscript $j$. 
\begin{equation}
    d\f^j(\lt) = a^j(\f,\lt) d\lt + \sum_j b^{jk}(\f,\lt) dW^k.
\end{equation}
Its evolution is governed by a diffusion term conventionally denoted by $a^j(\f,\lt)$, which may depend on all other stochastic variables, as well as the Langevin time explicitly. We incorporate $N$ independent Wiener processes $dW^j$ which obey the standard relations 
\begin{equation}
\left\langle \int_0^{\lt} dW^j \right\rangle = 0\quad \textrm{and} \quad  \left\langle \int_0^{\lt} dW^j  \int_0^{\lt} dW^k \right\rangle = \delta^{jk}\int_0^{\lt}  d\lt'.
\end{equation}
They affect the dynamical degrees of freedom $\f^j$ via the mixing matrix $b(\f,\lt)$. This general non-constant noise coefficient matrix may have a non-trivial dependence on Langevin time and the stochastic variables.

In the concrete case of stochastic quantization, we replace the discrete parameter $j$ with the combination of an discrete index for field degrees per spacetime point and the continuous parameter $x$. Hence the sum over $j$ turns into a combined sum and integral $\sum_j \rightarrow \sum_j \int dx$. Kronecker deltas remain for discrete indices, while we have to introduce Delta functions for spacetime $\delta_{jk} \rightarrow \delta_{jk}\delta(x-x')$. This leads us to the expression
\begin{equation}\label{eq:SDE_continuous_x}
\begin{aligned}
    & d\f^j(x,\lt) = a^j(\f,x,\lt) d\lt + \int dx' \sum_k b^{jk}(\f,x,x',\lt) dW^k(x',\lt)  ,
\end{aligned}
\end{equation}
which can be matched to our complex Langevin \cref{eq:CLE} using
\begin{equation}
    a^j(\f,x,\lt) = i\frac{\delta S[\f]}{\delta \phi^j(x)}, \quad b^{jk}(\f,x,\lt) = \sqrt{2} \delta^{jk}\delta(x-x').
\end{equation}
Note that in its standard form, the CL noise coefficients are constant in $\f$ and $\lt$.

As a first step, we need to take care of the drift term, which originates from a discretized action. To this end, one discretizes physical spacetime on which the field degrees live, turning continuous $x$ into a discrete set of coordinates $x_m$. The integral in the classical action may be approximated using a Newton-Cotes formula with the weights $\omega_m=\omega(x_m)$, which yields the following drift and noise terms for the continuous Langevin time SDE
\begin{equation}
    a^{j,m}(\f,\lt) = \frac{i}{\omega_m}\frac{\partial S[\phi]}{\partial \phi^j(x_m)}, \quad b^{jk,ml}(\f,\lt) = \sqrt{\frac{2}{\omega_m}} \delta_{jk}\delta_{ml}.
\end{equation}
This expression for the diffusion term $a$ and the noise coefficient matrix $b$ can be used straightforwardly in numerical schemes for stochastic differential equations. In the remainder of this section, we discuss some of these schemes in more detail.

The most common implementation of the CL dynamics deploys the simple Euler-Maruyama (EM) scheme. Discretizing Langevin time in equidistant steps of size $\Delta\lt$ we introduce the discrete Wiener process increment $\Delta W^j_\lambda=W^j_{\lambda+1}-W^j_{\lambda}$. The subscript $\lambda$ denotes at which Langevin time step the random variables are evaluated. The update step is then given by the following expression, where summation over repeated indices is implied
\begin{align}
    \nonumber &\f^{j,n}_{\lambda+1} = \f^{j,n}_{\lambda} + \Delta\lt \; \left[ \theta a^{j,n}\left(\f_{\lambda+1}\right) + (1-\theta)a^{j,n}\left(\f_{\lambda}\right)\right]  + b^{jk,nm} (\f_\lambda) \; \Delta W^{k,m}_\lambda \label{eq:EMscheme}   \\
   & \textrm{with} \quad \langle \Delta W^{j,n}_\lambda \rangle =0, \quad  \langle \Delta W^{j,n}_\lambda \Delta W^{k,m}_\lambda \rangle = \Delta\lt \delta_{jk}\delta_{nm}.
\end{align}
Here we actually refer to a whole class of EM schemes, which differ by the choice of a single real-valued parameter $\theta$. It controls the level of implicitness. For $\theta=0$ one recovers the fully explicit forward EM scheme, while for $\theta=1$ the implicit variant ensues. The choice of $\theta=1/2$ is special, as it refers to a semi-implicit Crank-Nicholson-like implementation of the EM scheme. 

In contrast to the deterministic Euler schemes, the different variants of the EM scheme for a non-trivial noise term share a numerical accuracy of strong order $\mathcal{O}(\sqrt{\Delta\lt})$. I.e. in general they perform worse than in the deterministic case, which is a common ailment afflicting the direct application of deterministic schemes to SDEs. In the case of simple CL dynamics we are fortunate however, in that the noise term remains trivial and thus the simple EM scheme can be shown to be of strong order $\mathcal{O}(\Delta\lt)$.

While the same numerical accuracy is shared among the different members of the EM family of schemes, their numerical stability varies significantly. It is well known that the forward EM scheme is at best conditionally stable, while the fully implicit scheme $\theta=1$ is robust against instabilities, being what is called in the literature L-stable. Similarly, it can be shown (c.f. Crank-Nicolson) that the semi-implicit scheme $\theta=1/2$ is also unconditionally asymptotically stable \cite{MILOSEVIC2013887}.

We stress that stability and accuracy are two separate qualities of a scheme, where stability only refers to the ability of the numerical solver to follow the true solution within the limitations placed by the accuracy of the scheme. Unconditional stability however also guarantees that as long as the true solution remains bounded, the scheme will not produce divergent runaway solutions. This property is what leads us to propose the deployment of (semi-)implicit solvers for complex Langevin, as it allows us to disentangle possible breakdown of the stochastic quantization prescription from a breakdown of the numerical solver.

In general, an implicit scheme is more costly than its explicit cousin at each individual update step. We need to solve a non-linear system of equations arising from the drift term $a^{j,n}(\f_{\lambda+1})$ in \cref{eq:EMscheme}, which is commonly implemented by a variant of Newton's method. As we will see, the favorable stability properties may however allow one to choose larger step sizes in Langevin time, leading to an overall reduction in computation cost.

Similarly as for deterministic differential equations, one may improve on the simple Euler schemes by developing \textit{Runge-Kutta} solvers for SDEs, usually referred to as \textit{SRK} schemes. They offer a straightforward way to increase the accuracy of the solver by combining approximations of the stochastic variables at intermediate steps within one update interval. The simplest of these is the \textit{Runge-Kutta Milstein} scheme \cite{KloedenPlaten,robler2010} of strong order ${\cal O}(\Delta\lt)$ for diagonal noise  
\begin{equation}\label{eq:RKMilScheme}
\begin{aligned}
    \f^{j,n}_{\lambda+1} = &\f^{j,n}_{\lambda} + \Delta\lt \; \left[ \theta a^{j,n}\left(\f_{\lambda+1}\right) + (1-\theta)a^{j,n}\left(\f_{\lambda}\right) \right] + b^{j,n} \left(\f_{\lambda}\right) \; \Delta W^{j,n}_{\lambda} \\ 
              &+ \frac{1}{2\sqrt{\Delta\lt}}\left( b^{j,n}(\Upsilon_\lambda ) - b^{j,n}(\f_{\lambda}) \right)\left\{ (dW^{j,n}_\lambda)^2 - \Delta\lt \right\} \quad \textrm{with} \\
    \Upsilon_\lambda^{j,n} = & \f_{\lambda}^{j,n} + a^{j,n}\left(\f_{\lambda}\right)\Delta\lt + b^{j,n} \left(\f_{\lambda}\right) \sqrt{\Delta\lt}.
\end{aligned}
\end{equation}
Again we have indicated a whole family of schemes, whose implicitness is governed by the $\theta$ parameter. Note that these schemes differ from a naive application of the deterministic second-order Runge-Kutta (RK2) prescription through the presence of a term quadratic in the Wiener process in the second line of \cref{eq:RKMilScheme}. In case of simple CL \cref{eq:CLE} with constant real noise coefficients, the Milstein scheme reduces to the EM scheme, reaffirming that for trivial noise the simplest algorithm already offers order 1.0 accuracy.

Let us also touch on higher-order schemes. The next order one can reach is 1.5 \cite{KloedenPlaten}, at which the SRK prescription for constant additive noise reads
\begin{align}\label{eq:SRK15}
    \f_{\lambda+1} =& \f_\lambda + \Delta\lt \; \frac12 \left[a\left(\f_{\lambda+1}\right) + a\left(\f_{\lambda}\right) \right] \\ 
    \nonumber& +  b \; dW_\lambda + \frac{1}{2\sqrt{\Delta\lt}} \left\{ a(\Upsilon_+^\lambda) - a(\Upsilon_-^\lambda) \right\}\left\{ dZ_\lambda - \frac12 dW_\lambda d\lt \right\}  \\
    \nonumber \textrm{with} & \quad \Upsilon_\pm^{\lambda} =  \f_\lambda + a\left(\f_\lambda\right)\Delta\lt \pm b \sqrt{\Delta\lt}.
\end{align}
Here we use vector notation and we have explicitly chosen $\theta=1/2$ for simplicity of the presentation. In this equation, we find that a genuinely new contribution arises even for trivial noise. It consists of a combination of the drift term together with Wiener processes $dW$ and $dZ$. The latter one refers to additional independent processes with the same mean and variance as the $dW$ as well as $\langle dW^j dZ^j \rangle = 0$. By incorporating the proper contributions arising from Ito's lemma in the series expansions underlying these Runge-Kutta schemes one may thus construct consecutive improvements to the naive EM scheme.

In our study, we will draw upon the implementation of the above-mentioned schemes through the SDE module in the \textit{DifferentialEquations.jl} \cite{DifferentialEquations.jl-2017,rackauckas_stability-optimized_2018} library provided in the Julia language. The concrete implementations of \cref{eq:EMscheme,eq:RKMilScheme,eq:SRK15} differ slightly due to performance improvements outlined in the literature (see the documentation of \cite{DifferentialEquations.jl-2017}), which however has no effect on their stability and accuracy properties.

All the methods listed above can be implemented with an adaptive step-size prescription. This offers two concrete benefits. On the one hand, the stability properties of a simulation can be improved, as the step size is adapted to fulfill the Courant–Friedrichs–Lewy stability condition at each update. For stiff problems and explicit solvers, this approach is limited in practice by the step size becoming so small that the number of steps along Langevin time grows beyond available computational power. On the other hand adaptive step size also allows us to increase the step size at intermediate times to reduce the computational burden while staying within a predefined accuracy tolerance for the update step. One drawback of adaptive step size is the fact that an analytic investigation of the properties of the solver becomes more involved. We will thus deploy adaptive step size in all simulations except those where we study the finite time discretization artifacts and look at the corrections from the discretized Fokker-Planck equation. 

Many different adaptive step prescriptions are deployed in the literature. One of the more sophisticated approaches implemented e.g. in the Julia library compares updates of solvers of different order and takes the difference as an error estimate\cite{rackauckas2017adaptive}. The step size then is chosen to keep this error estimate below a pre-defined threshold. More simply we may monitor the size of the drift term in the Langevin equation and adjust the time step such that the change induced by the drift term remains below a certain threshold. We have found that some adaptive step-size algorithms implemented in the literature do not contain a limit on the maximum step size, which may spoil the accuracy of the outcome and required us to implement such an upper limit by hand.

\subsection{On the issue of large excursions}
\label{sec:excusions}

Having reviewed different explicit and implicit prescriptions for the solution of the complex Langevin SDE, we may now explore how these methods fare in addressing the issue of stability. A common challenge that plagues complex Langevin simulations is the occurrence of large excursions. While the overwhelming majority of trajectories contributing to the final expectation value are located in a well-contained area around the origin, some paths are found to venture significantly further out into the complex plane. A simple example of this behavior can be found in the system of a single degree of freedom, evolving in the potential $V(\phi) = i\phi^4$. On average it leads to paths that stay within around 2 dimensionless units from the origin, however excursions up to $|\phi| \sim 10$ sporadically occur. 

\begin{figure}
         \includegraphics[scale=0.35]{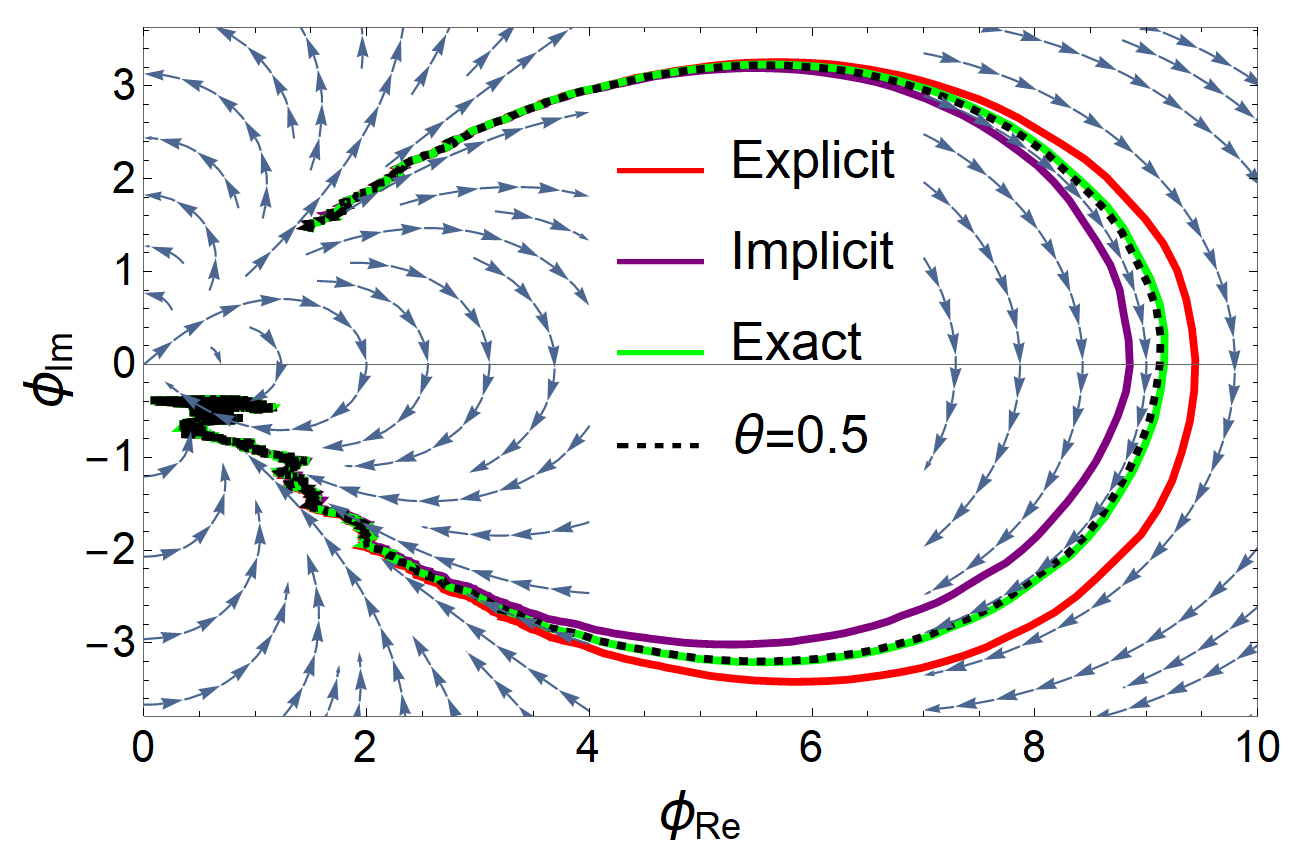}
    \caption{\label{fig:Explicit_Vs_Implicit} Example of complex Langevin paths for a single degree of freedom in the potential $V(\phi) = i\phi^4$ based on different solvers: (red) explicit EM, (violet) implicit EM, (dashed) unitary $\theta=1/2$ EM and the exact solution given as green line. Each path is initiated at $(1.5,1.47)$ close to the divergent flow line. The Langevin time step size is kept constant at $\Delta \lt=10^{-4}$. Note the characteristic over and undershooting of the explicit and implicit method respectively.}\label{fig:FlowPhi4Mod}
\end{figure}

These excursions can be understood by inspecting the flow field $-4i \phi ^3$, as shown in \cref{fig:FlowPhi4Mod} along the line where $\phi _{R} = \phi _{I}$. As $\phi _{I} \gtrsim \phi _{R}$ the flow lines tilt upward, while for $\phi _{I} \lesssim \phi _{R}$ they tilt downwards. As one moves exactly on top of the line, the field lines will keep going straight out towards infinity. This is a property of the continuum theory and not an artifact of the numerical solution.

In principle, this is not a problem, since the noise term makes sure that one never stays on this line indefinitely. However, as the size of $\phi$ increases, the size of the flow $4i \phi^3$ also increases significantly compared to the noise term. In turn, after the noise kicks the system away from the diverging path, it now follows a path dominated by the drift term with only a small contribution of the noise term. Such a path tends to go out to even larger values of $|\phi|$ until it eventually returns to the dominating region. A representative example is shown in \cref{fig:Explicit_Vs_Implicit} as the green solid line.

It has been understood that such excursions constitute one of the reasons for the stability issues of numerical implementations of complex Langevin. Let us have a look at how well the true path is recovered by the simple EM schemes for different settings of implicitness for a fixed step size. In general, the explicit method is prone to overshooting the correct trajectory (red solid line), while the fully implicit method also fails to stay close to it but does so by undershooting the correct result (violet solid line). The overshooting of the explicit method easily leads to divergent behavior as the errors accumulate. For the implicit method on the other hand the simulation remains stable. Note however that its accuracy still suffers due to the deviation from the true trajectory. The semi-implicit EM scheme on the other hand combines the best of both worlds, as it offers the unconditional stability of the implicit scheme and limits the undershooting to a minimum, as can be seen in the dashed black line in \cref{fig:Explicit_Vs_Implicit}. Note that due to the large flow the actual Langevin time spend in one of these excursions is very small compared to the total length of the trajectories needed to accumulate reasonable statistical uncertainties.

We can understand the behavior seen in \cref{fig:Explicit_Vs_Implicit} already from an inspection of the discretization prescription in the free theory. There the drift term is linear ($a(\phi) = iM\phi$) and we can rewrite the EM scheme of \cref{eq:EMscheme} as
\begin{equation}
    \phi^{\lambda+1} = \left( 1 - i \Delta \lt \theta M \right)^{-1} \left\{ (1 + i \Delta \lt(1-\theta)M)\phi^{\lambda} + \sqrt{\Delta \lt}\eta^\lambda \right\}.
\end{equation}
Taking the expectation value of the field at $\lt=(\lambda + 1)\Delta \lt$ we get
\begin{equation}\label{eq:expectationvalue_discField}
   | \langle  \phi^{\lambda+1}  \rangle | = \frac{ |1 + i \Delta \lt(1-\theta)M|}{\left| 1 - i \Delta \lt \theta M \right|} | \langle \phi^{\lambda} \rangle |.
\end{equation}
For $\theta = 0$ the fraction simplifies to $\left| 1 + i \Delta \lt M \right| > 1$, which induces an increase in the magnitude of the field value for each step. On the other hand for $\theta=1$ one finds $\left| 1 - i \Delta \lt M \right|^{-1} < 1$, which represents shrinkage of the magnitude. For the special case of  ($\theta=\frac12$) we obtain a semi-implicit scheme with $|\langle \phi^{\lambda+1} \rangle | =\left| 1 + i\Delta \lt M \right|^{-1} \left| 1 - i \Delta \lt M \right|  |\langle \phi^{\lambda} \rangle | = | \langle \phi^{\lambda} \rangle |$ which in the free case exactly preserves the magnitude of the expectation value of the field. 

We chose the above example to illustrate a key qualitative difference between solvers of varying degrees of implicitness. It should be mentioned that for the specific scenario shown here, the differences are only sizeable since we do not use an adaptive step size. A $\Delta \tau _L =10^{-4}$ allows for stable dynamics close to the origin where the drift is small. However at $|\phi | \sim 10$ we have a relatively large drift term of around $\Delta \tau _L 4 \phi ^3 = 0.4$. In practice, using adaptive step size one would reduce $\Delta \lt$ along the excursion, keeping the deviation from the exact solution small. The conclusion that explicit schemes accumulate errors according to overshooting of the true trajectory and implicit schemes according to undershooting it remains unchanged.

\section{Towards stable real-time simulations of the quantum anharmonic oscillator}
\label{sec:towards}

\subsection{Formulating and discretizing the model}
\label{sec:model}

Our main goal in this study is to implement stable simulations of the early real-time dynamics of a strongly coupled quantum anharmonic oscillator. This system amounts to a (0+1)d field theory prototype and has been studied in the literature in detail \cite{BergesSexty2007,Alexandru:2016gsd}, establishing itself as a benchmark for the success of different real-time approaches. Real-time expectation values for an observable $\mathcal{O}$ arising in a system that evolves from a mixed initial state $\rho$ can be described via the Schwinger-Keldysh closed time-path formalism. In its canonical implementation, it deals with field degrees of freedom placed on a time contour, with both a forward-facing branch along the time axis (housing $\phi_+$) and a backward branch (housing $\phi_-$), both of which are attached at the initial time to the density matrix $\rho(\phi_1,\phi_2)$
\begin{equation}
    \langle \mathcal{O}(\phi) \rangle = \frac{1}{Z} \int d\phi_1 \int d\phi_2 \; \rho(\phi_1,\phi_2) \int_{\phi_2}^{\phi_1} D\phi^+ D\phi^- \; \mathcal{O}(\phi) \; e^{iS[\phi_+] - iS[\phi_-]}.
\end{equation}
Let us consider the case of thermal equilibrium with inverse temperature $\beta=1/T$. The density matrix takes on the standard Boltzmann form $\rho\propto {\rm exp}[-\beta H]$ and we may conveniently absorb the sampling over initial conditions into a path integral along the imaginary time axis, compactified to a length of $\beta$
\begin{equation}
    \langle \mathcal{O}(\phi) \rangle = \frac{1}{Z} \int D\phi_E e^{-S_E[\phi_E]} \int_{\phi_E(\beta)}^{\phi_E(0)} D\phi^+ D\phi^- \; \mathcal{O}(\phi) \; e^{iS[\phi_+] - iS[\phi_-]}.
\end{equation}
The corresponding Schwinger Keldysh contour now contains three parts: the forward and backward real-time branch, as well as the Euclidean contour all of which contribute with their own action and which we will summarize as $iS[\phi_+] - iS[\phi_-] - S_E \rightarrow iS[\phi]$.

The real-time action for the anharmonic oscillator explicitly reads 
\begin{equation}
    S = \int d\rt \left\{ \frac{1}{2}  \left( \frac{\partial \f}{\partial \rt} \right)^2 - V(\f) \right\}, \quad V(\f) = \frac{1}{2}m \f^2 + \frac{\lambda}{4!}\f^4,
\end{equation}
where $\lambda$ refers to the coupling constant. We give all of our results in units of $m$, which is the same as setting $m=1$. 
The first step to take is to discretize the time coordinate $x_0\in\mathbb{C}$ along the Schwinger-Keldysh contour. To this end, we introduce a contour parameter $\xi\in\mathbb{R}$ which on the forward and backward branch refers to a real-valued time, while on the Euclidean branch refers to a negative imaginary time. The time integral in the action becomes a line integral over the contour parameter, discretized into $N_{\cal C}$ steps $a_j$. The fields, evaluated at the discrete real-time steps are denoted by $\phi(\sum_j a_j)=\phi_j$. We deploy the trapezoidal rule for the action integral, which amounts to averaging over the left and right Riemann sums. Consistently we approximate the derivative by finite differences choosing forward difference at the point $j$ and backward differences at the point $j+1$, leading us to the standard expression
\begin{equation}\label{eq:discretized_action}
    S = \frac{1}{2}\sum_j \left\{  \frac{ \left(\f_{j+1} - \f_j\right)^2}{a_j} - a_j \left[ V(\f_{j+1}) + V(\f_j)  \right] \right\}.
\end{equation}
To implement the complex Langevin equations of motion, we need to calculate the drift term  $i\frac{\delta S}{\delta \phi_j}$. Using the discretization scheme above we obtain
\begin{align}
        \label{eq:discretized_action_derivaitve}
    i\frac{\delta S[\phi]}{\delta \phi_j} =  \frac{i}{\frac{1}{2}\left(|a_{j}| + |a_{j-1}|\right)}\Big\{ &\frac{\phi_j - \phi_{j-1}}{a_{j-1}}- \frac{\phi_{j+1} - \phi_j}{a_j}
    - \frac12 \left[a_{j-1} + a_j\right] \frac{\partial V(\phi_j)}{\partial \phi_j} \Big\}.
\end{align}
The factor $\frac{1}{\frac{1}{2}\left(|a_{j}| + |a_{j-1}|\right)}$, which we have included here explicitly in the drift term must then also be consistently included in the noise term $b^{jk} = \sqrt{ \frac{2}{\frac{1}{2}\left(|a_{k}| + |a_{k-1}|\right)}}\delta_{jk}$. Note however that via a rescaling of the drift term, one may drop the factors of $a$ if one consistently drops them also from all the Kronecker deltas associate with functional derivatives. At this stage we have only discretized the physical coordinates, leaving us with a continuous Langevin time prescription to stochastically quantize the anharmonic oscillator.

The discretization of the action already introduces numerical artifacts in the solution of the continuous-time complex Langevin equation. In order to make sure that we do not misinterpret such errors as arising from the finite Langevin time discretization in an actual simulation, we take a closer look at them here.

Analogous to constructing the transfer matrix operator we can go backward from the discretized path integral to the corresponding operator expressions while leaving the real-time step size $a_j$ finite. In that case, we have to deal with the fact that the Campbell-Baker-Hausdorff formula gives non-trivial contributions when decomposing the action integral into individual exponentials. Let us define the exponentiated Hamiltonian of the system via the following matrix elements
\begin{align}
    \langle \phi_{j+1}| \exp(i a_j H ) | \phi_j\rangle = {\rm exp}\left[\frac12  \frac{ \left(\f_{j+1} - \f_j\right)^2}{a_j} -  \frac{1}{2}a_j V(\f_j)-  \frac{1}{2}a_j V(\f_j) \right].
\end{align}
According to our \cref{eq:discretized_action} the RHS contains the potential evaluated at neighboring values of the field $\phi_{j+1}$ and $\phi_{j}$, which requires that the potential operator acts on both the left and right state. Since only one complete set of momentum eigenstates is involved in transforming the kinetic term back to its operator form we end up with the expression
\begin{align}
    \exp(i a_j \hat H_{\rm s} ) =&  {\rm exp}\left[  \frac{ a_j }{2} V(\hat\f)\right]  {\rm exp}\left[  \frac{ a_j \hat \pi^2}{2 } \right] {\rm exp}\left[  \frac{ a_j }{2} V(\hat\f)\right] +{\cal O}(a^2).\label{eq:half_dis}
\end{align}
Had we considered just one single potential term in \cref{eq:discretized_action} the corresponding operator expressions would have turned out to be 
\begin{align}
        \exp(i a_j \hat H_{\rm r} ) = &   {\rm exp}\left[  \frac{ a_j \hat \pi ^2}{2 } \right] {\rm exp}\left[   a_j  V(\hat\f)\right] +{\cal O}(a) \quad {\rm or} \label{eq:left_dis}\\
    \exp(i a_j \hat H_{\rm l} ) = & {\rm exp}\left[  a_j V(\hat\f)\right]  {\rm exp}\left[  \frac{ a_j \hat \pi ^2}{2 } \right] +{\cal O}(a). \label{eq:right_dis}
\end{align}

Using as parameters $\lambda=24$ and $m=1$, similar to what we will deploy in the actual complex Langevin simulations in the following sections, we can now study the effects of the finite real-time spacing explicitly. To this end we compute the forward correlator $\langle \phi(x_0)\phi(0)\rangle$ using matrix mechanics in the truncated Hilbert space spanned by the 32 lowest-lying energy eigenstates of the harmonic oscillator, according to the different effective Hamilton operators $\hat H$ defined above.

\begin{figure}   
    \includegraphics[scale=0.45]{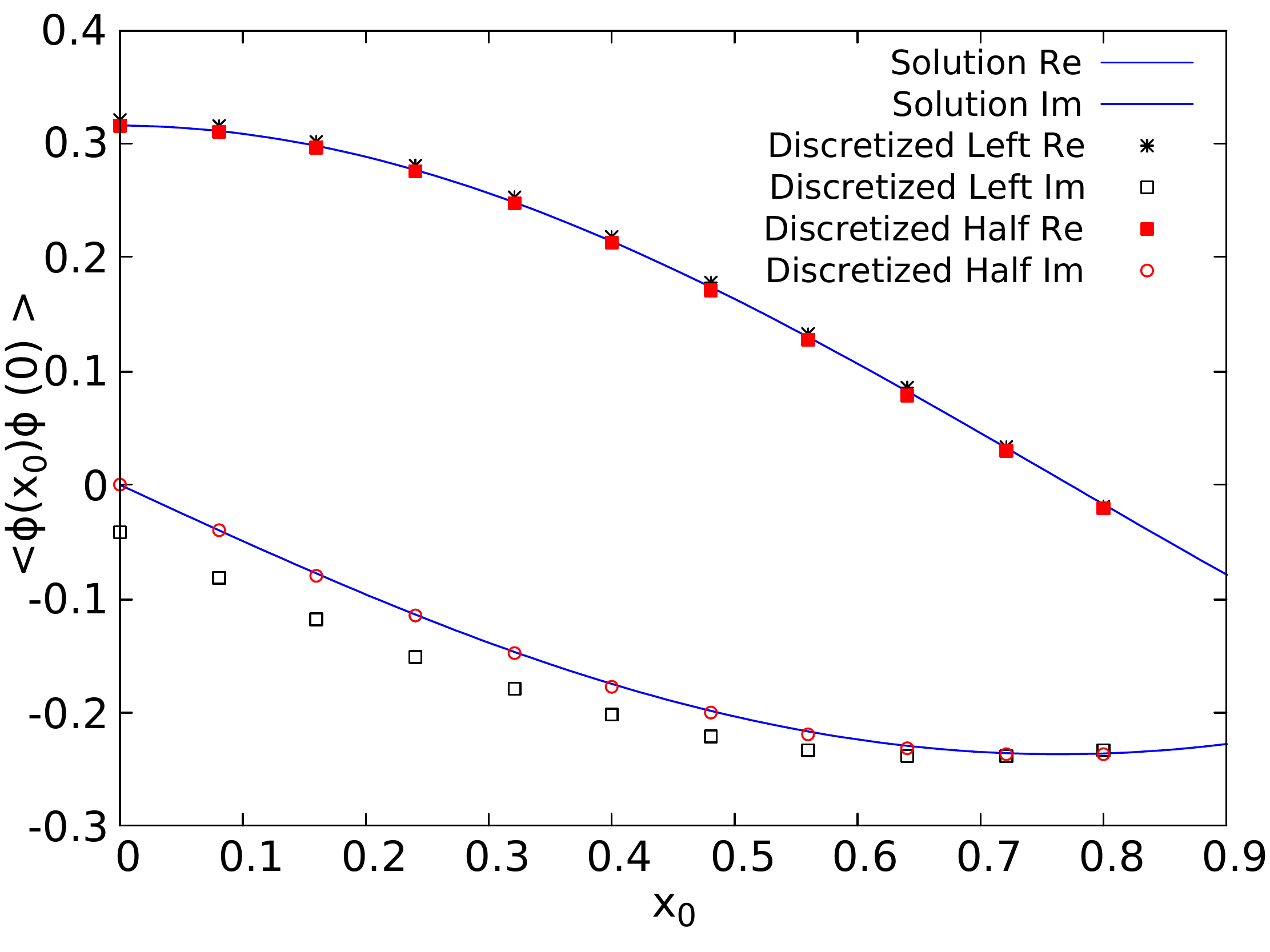}
    \caption{Visualization of real-time discretization artifacts in the unequal-time correlation function $\langle \phi(x_0)\phi(0)\rangle$. True values (solid lines) are obtained from matrix mechanics in the truncated Hilbert space spanned by 32 energy eigenstates of the harmonic oscillator. Note the shift in the imaginary part when using the first order discretization of \cref{eq:left_dis} (open squares) for a lattice spacing of $am=0.08$. The symmetric discretization of \cref{eq:half_dis} significantly improves the agreement with the continuum results as seen in the open circles.}\label{fig:DisEffect}
\end{figure}

As can be seen in \cref{fig:DisEffect}, we find a characteristic artifact introduced by the finite real-time steps. Its main manifestation is the appearance of a non-zero value of the imaginary part of the correlator at the origin. Neighboring values of the imaginary part are correspondingly also shifted away from their true values. When using the ${\cal O}(a)$ discretization of \cref{eq:left_dis} this effect is of the order of $12\%$ for $a=0.08$ (with respect to the maximum value the imaginary part of the correlator takes on.) The strength of the effect scales as expected linearly with $a$ so that we obtain $6\%$ deviation for $a=0.04$ and $4\%$ deviation for $a=0.08/3$. Switching from \cref{eq:left_dis} to (\ref{eq:right_dis}) changes the sign in the shift of the imaginary part, indicating that the discretization amounts to a complex phase factor. Combining the opposing phase factors in the symmetric formulation of \cref{eq:half_dis} cancels out the effect to a significant extent, reducing the deviation from the continuum results to $0.2\%$ already for $a=0.08$. We will make sure to keep these discretization artifacts below the percent level in the Complex Langevin simulations in the following.

\subsection{Regularizing the model}
\label{sec:regularizing_CLE}

To make the continuum theory well-defined on the canonical Schwinger-Keldysh real-time contour, we need to introduce an infinitesimal damping term into the otherwise purely oscillatory behavior of the Feynman path integral weight. Otherwise, the continuous Langevin-time evolution will not be able to converge to a finite result. In an analytic setting, this is conventionally achieved by introducing by hand an additional term $R>0$ in the action such that the new effective action reads $\bar S = S + iR(\f,\epsilon)$. A simple example of such a regulator is $R = \frac12 \epsilon \f^2$ is e.g. discussed in \cite{Nakazato1986}.

It is long known that such a regulator also controls the rate of convergence in a complex Langevin simulation \cite{BergesSexty2007, BergesIon2005}. Similar to an analytic computation where the correct result is obtained by setting $\epsilon\to 0$ only a posteriori, a simulation operates at finite $\epsilon$, which, depending on its chosen value may significantly distort the computed expectation values. Only after an extrapolation over several different simulations do we recover the true solution. It goes without saying that the closer we can simulate to the correct solution, the less troubled the extrapolation procedure will be. Therefore we will attempt to deploy as small a regulator as possible.

In this section, we will briefly discuss the classic approach to introduce a tilt in the Schwinger-Keldysh contour \cite{BergesSexty2007} and report on our observation that the implicit scheme itself offers a regularization by construction.

\subsubsection{Tilted Schwinger-Keldysh contour}\label{sec:TiltedSK}

One way to regularize our model is to introduce an imaginary tilt in the Schwinger-Keldysh contour, as deployed e.g. in ref.\cite{BergesSexty2007} and shown in the leftmost panel of \cref{fig:contours}. Since in the thermal setting we are interested in correlators on the forward contour (the values of the mixed correlators are related via the KMS relation) one tries to keep the tilt on the forward branch small. The backward branch on the other hand may tilt downward more steeply, as long as it reaches the negative imaginary axis before $-i\beta$. To be more concrete, the tilted Schwinger-Keldysh contour (fig. \ref{fig:contours} a) has two distinct parts. Part one (${\cal C}_1$) is tilted under an angle $\alpha$ from 0 to $t_{max} - {\rm sin}(\alpha)\beta i$. Part two (${\cal C}_2$) is tilted such that it arrives at the imaginary axis at the point $-i\beta$, which due to periodic boundary conditions coincides with the starting point of the first part of the contour.

The information about the shape of the real-time contour is fully contained in the choice of the, in general complex, time step $a_j$. It denotes the distance between point $j$ and $j+1$ on the contour and appears explicitly in the complex Langevin drift term in \cref{eq:discretized_action_derivaitve}. A tilt of the real-time contour manifests itself as a non-zero imaginary part in $a_j$ such that 
\begin{equation}
\begin{aligned}
    \bar S =& \frac12 \sum_j \left\{  \frac{ \left(\f_{j+1} - \f_j\right)^2}{a_j^R + ia_j^I} - (a_j^R + ia_j^I) \left[ V(\f_{j+1}) + V(\f_j)  \right] \right\} \\ 
    =& \frac12 \sum_j \left\{  \frac{ \left(\f_{j+1} - \f_j\right)^2}{|a_j|}(a_j^R - ia_j^I) - (a_j^R + ia_j^I) \left[ V(\f_{j+1}) + V(\f_j)  \right] \right\} \\
    =& S + \frac12\sum_{j} \left\{  \frac{ \left(\f_{j+1} - \f_j\right)^2}{|a_j|}( - ia_j^I) -  (ia_j^I) \left[ V(\f_{j+1}) + V(\f_j)  \right] \right\} \\
    =& S + \frac12 \sum_{j} \left\{  \frac{ \left(\f_{j+1} - \f_j\right)^2}{|a_j|} +  \left[ V(\f_{j+1}) + V(\f_j)  \right] \right\} ( - ia_j^I) \\
    =& S + i\sum_{j}R(\phi,a_{j}^{I}).\label{eq:tiltreg}
\end{aligned}
\end{equation}
We see that if $a_i$ has a negative imaginary part, the overall prefactor becomes $(-ia^I_j)=+i| \textrm{Im}(a_i)|$, turning the corresponding $R>0$ into a positive quantity. Coming to the conclusion that such a positive term $R$ successfully acts as a regulator however is not as straightforward as it appears at first sight. In the case of a complex Feynman weight, the field themselves becomes complexified and $R$ exhibits both a real- and imaginary part. In the free theory ref.\cite{Nakazato1986} has shown that a positive $R= \frac12\epsilon\phi^2$ allows us to take a well defined late Langevin-time limit of the complex Langevin dynamics with a regularization of the two-point function that amounts to $\langle \phi(k)\phi(-k) \rangle = i/(k^2-m^2+i\epsilon)$ supporting the downward tilt of the Schwinger-Keldysh contour as an appropriate regulator.

\begin{figure}
\begin{minipage}{0.3\linewidth}
\begin{tikzpicture}
\draw [->,thick,gray] (0,0) -- (3,0) node[above] {Re};
\draw [->,thick,gray] (0,0) -- (0,-3) node[left] {Im};

\draw [-,thick] (0,0.0) -- (2.5,-0.1);
\node[above] at (1.5,-0.05) {${\cal C}_1$};
\draw [-,thick] (2.5,-0.1) -- (0,-2.0);
\node[below] at (1.5,-1.0) {${\cal C}_2$};
\node[left] at (0,-2) {$-i\beta$};
\end{tikzpicture}
\centering
\textbf{a}
\end{minipage}
\begin{minipage}{0.3\linewidth}
\begin{tikzpicture}
\draw [->,thick,gray] (0,0) -- (3,0) node[above] {Re};
\draw [->,thick,gray] (0,0) -- (0,-3) node[left] {Im};

\draw [-,thick] (0,0.0) -- (2.5,0.0);
\node[above] at (1.5,-0.0) {$S_1$};
\draw [-,thick] (2.5,-0.0) -- (0,-2.0);
\node[below] at (1.5,-1.0) {$S_2$};
\node[left] at (0,-2) {$-i\beta$};
\end{tikzpicture}
\centering
\textbf{b}
\end{minipage}
\begin{minipage}{0.3\linewidth}
\begin{tikzpicture}
\draw [->,thick,gray] (0,0) -- (3,0) node[above] {Re};
\draw [->,thick,gray] (0,0) -- (0,-3) node[left] {Im};

\draw [-,thick] (0,0.0) -- (2.5,0.0);
\node[above] at (1.5,-0.0) {$S_1$};
\draw [-,thick] (2.5,0.0) -- (0,0.0);
\node[below] at (1.5,-0.0) {$S_2$};
\draw [-,thick] (0,0.0) -- (0,-2) node[left] {$-i\beta$};
\node[left] at (0,-1.0) {$S_E$};
\end{tikzpicture}
\centering
\textbf{c}
\end{minipage}
\caption{Three different realizations of the thermal Schwinger-Keldysh contour for a system with temperature $T=\frac1\beta$. The leftmost setting (a) corresponds to the contour adopted e.g. in ref.\cite{BergesSexty2007}. Our first goal is to be able to remove the tilt on the forward contour (b) and ultimately in preparation for the non-equilibrium setting to move both the forward and backward branch very close to the real-time axis (c). We find that the inherent regularization of the implicit solver allows us to realize scenarios (b) and (c) in practice.}
\label{fig:contours}
\end{figure}

Reducing the tilt of ${\cal C}_1$ reduces the strength of the regularization. It is well known and we have reconfirmed in our numerical experiments that concurrently the stochastic dynamics become more and more stiff. I.e. when deploying an explicit scheme the probability to encounter runaway solutions increases significantly as the tilt is reduced. In the classic work of \cite{BergesSexty2007}, an explicit solver was combined with a $0.01\beta$ tilt of the contour. While this tilt is small at the early times considered in previous and also this study, it already introduces a deviation from the true solution in the unequal-time correlation functions, which goes beyond the statistical errorbars of the simulation. With the goal of extending CL simulations to later real-times in the future, we will be urged to reduce the tilt even further.

As an example let us carry out a simulation using a similar setup as in \cite{BergesSexty2007}, with a tilt of $0.01\beta$ in the anharmonic oscillator action (\cref{eq:discretized_action}). In order to remain in the region where complex Langevin converges to the correct result, we select as maximum real-time extent $x_0^{\rm max}=0.5$. As a solver the general Euler-Maruyama scheme (\cref{eq:EMscheme}) with adaptive step-size\footnote{All simulations based on the implicit scheme can be carried out without adaptive step-size. The numerical cost in that case will simply be higher, as an overall smaller step-size is needed to reach the same accuracy. Nevertheless, compared to simulating with an explicit scheme, we can deploy a much larger step size. The implicit scheme already works well with $\Delta\lt=10^{-3}$, while the explicit scheme requires us to go to $\Delta\lt=10^{-5}$.} is chosen. The $\theta$ value in \ref{eq:EMscheme} is set to $\theta=\frac12$ corresponding to a semi-implicit scheme, which, as we have seen in section \ref{sec:excusions}, preserves the magnitude of the expectation value throughout the simulation well. We average over a total of 500 trajectories, each of which reaches a total Langevin time of $\lt m=100$. Observables are read out every $\delta \lt m = 0.1$ in Langevin time. 

\begin{figure}\centering
        \includegraphics[scale=0.42]{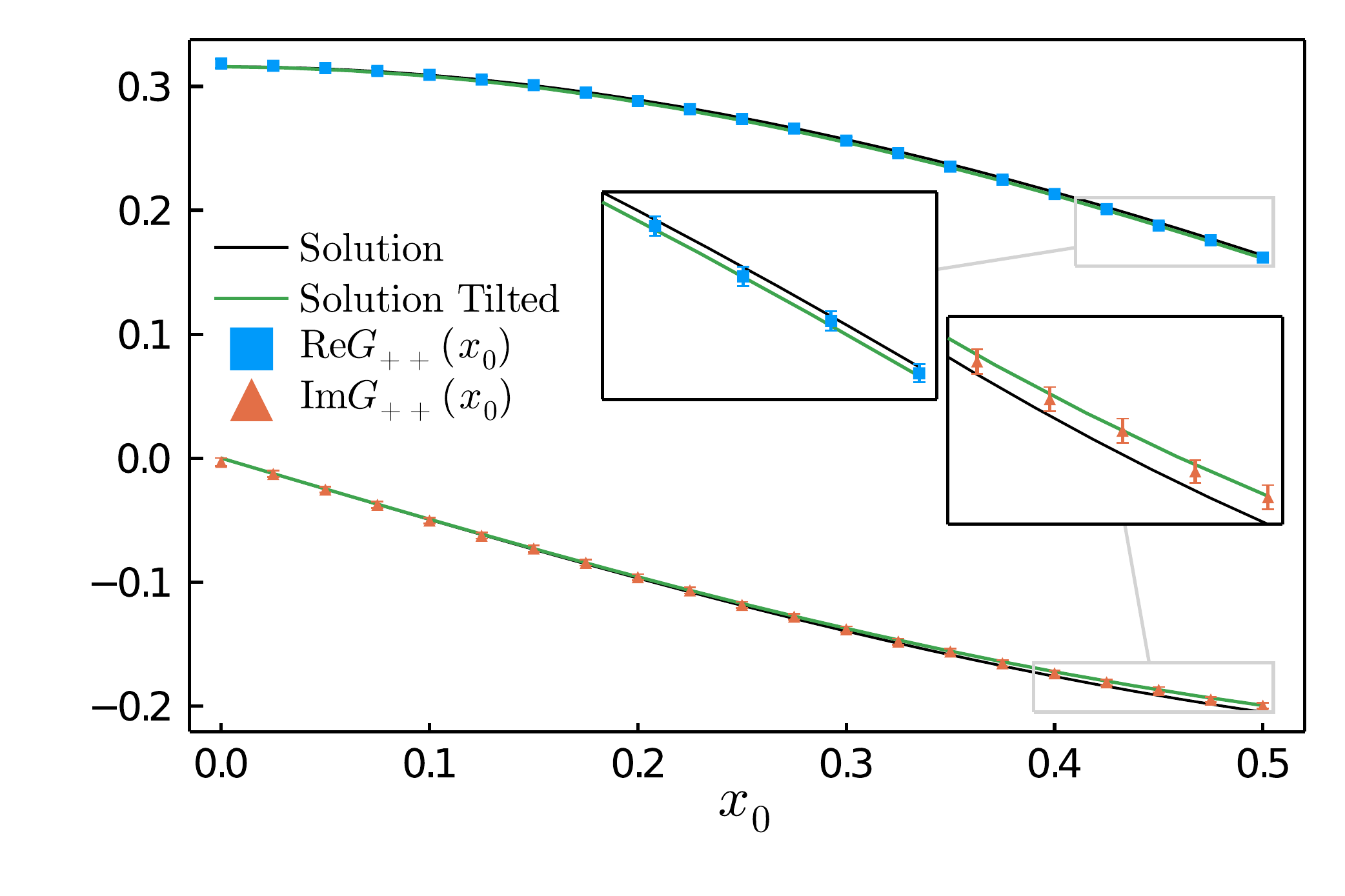}
        \includegraphics[scale=0.42]{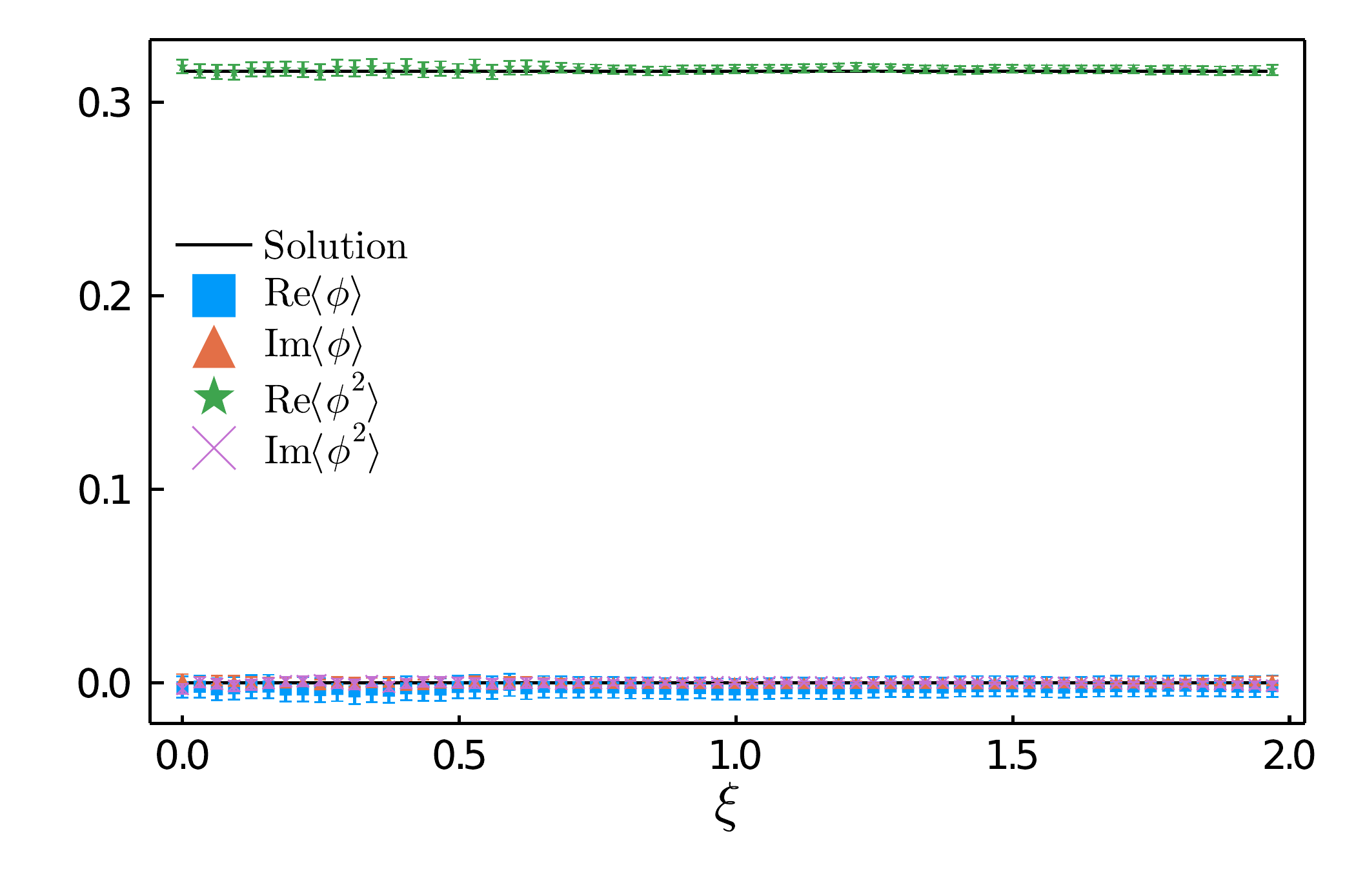}
    \caption{Simulation results for the anharmonic oscillator based on the semi-implicit Euler-Maruyama scheme on a $0.01\beta$ tilted contour: (top) unequal-time correlation function $G_{++}(x_0)=\langle \phi(0)\phi(x_0)\rangle- \langle \phi(0)\rangle\langle\phi(x_0)\rangle$ plotted for real-time values along the forward branch of the tilted Schwinger-Keldysh contour together with the true solution on the real-time axis (black solid) and along the tilted contour (green solid). Close to $x_0 =0.5$ our simulation can already distinguish between the two. (bottom) The field expectation value $\langle \phi\rangle$ and the equal-time correlation function $\langle\phi^2\rangle$ evaluated on both branches. Agreement with the true solution is observed within errors within uncertainty (errors appear larger here only since the y-axis scale is reduced compared to the top panel).}
    \label{fig:tilted_contour_two_point_fuction}
\end{figure}

We plot the real- and imaginary part of the unequal time correlation function $G_{++}(x_0)=\langle \phi(0)\phi(x_0)\rangle-\langle \phi(0)\rangle\langle\phi(x_0)\rangle$ on the forward branch vs. the contour parameter $\xi$ in the top panel of \cref{fig:tilted_contour_two_point_fuction}. While a small effect, we can already distinguish between the analytic solution on the real-time axis (black solid) and the solution on the tilted contour ( green solid ) within the precision of our simulation. The analytic solution here is obtained again using matrix mechanics in the truncated Hilbert space spanned by the 32 energy eigenstates of the harmonic oscillator.

As shown in the magnified insets, close to $x_0 =0.5$ the tilt leads to a visible deviation from the true solution. I.e. such a tilt does affect the solution at early times and will become sizeable once the simulation can be extended to a phenomenologically relevant real-time extent.

In the lower panel of \cref{fig:tilted_contour_two_point_fuction} the field expectation value $\langle \phi \rangle$ and the equal time correlation function $\langle \phi^2\rangle$ are plotted vs. the contour parameter along both branches of the contour. We find that they agree with the constant value predicted by the true solution. I.e. as is known, these quantities are less susceptible to the tilt as the unequal-time correlation function. 

\subsubsection{Regularization via an implicit scheme}
\label{sec:implreg}

Previous studies and the preceding subsection have shown that introducing a large enough tilt in the Schwinger-Keldysh contour, as in \cref{sec:TiltedSK}, allows us to regularize the oscillatory behavior of the path integral. Depending on the size of the tilt it does so effectively enough for even an explicit solver to capture the ensuing complex Langevin dynamics. The price to pay is a systematic deviation of the correlation function from the result on the real-time axis which grows with the maximum extent of the forward contour.

In this study, our goal is to explore the potential of implicit solvers for complex Langevin. We have already seen how their simplest formulation, in form of the EM scheme, avoids the occurrence of runaway solutions in \cref{sec:excusions}. We will return to the evolution equations and show that the formulation of the general EM scheme harbors additional terms, which play a role in the regularization of the path integral itself. 

Let us focus on the simple but relevant case of the free theory here with $V(\phi) = \frac12 m^2\phi^2$. The update step of the general EM scheme reads
\begin{equation}\label{eq:EMSchemeSQM}
    \phi_j^{\lambda + 1} = \phi_j^{\lambda} + i \epsilon_j\left[ \theta \frac{\partial S^{\lambda +1}}{\partial \phi_j} + (1-\theta)\frac{\partial S^{\lambda}}{\partial \phi_j}  \right] + \sqrt{\epsilon_j}\eta_j^\lambda,
\end{equation}
where $\epsilon_j=\frac{\Delta \lt}{|\omega_j|}$. To simplify the derivation below, we assume without loss of generality that the step size $\epsilon_i = \epsilon$ is constant along the contour. In the free theory we may in addition write the action in a simple matrix form as $\frac{\partial S^{\lambda}}{\partial \phi_j} = M\phi^\lambda_j$. Substituting this into \cref{eq:EMSchemeSQM} yields
\begin{equation}
    \left( I - i \epsilon\theta M \right)\phi^{\lambda+1} =  \left\{ (I + i \epsilon(1-\theta)M)\phi^{\lambda} + \sqrt{\epsilon}\eta^\lambda \right\}.
\end{equation}
The explicit entries in $M_{ij}$ are obtained via \cref{eq:discretized_action_derivaitve} as
\begin{equation}
    M_{jk} = 
\begin{cases}    
\frac{1}{a_{j-1}} + \frac{1}{a_j} - \frac12 \left[a_{j-1} + a_j\right]m^2,& j=k\\
 -\frac{1}{a_j},& j = k-1 \\
-\frac{1}{a_{j-1}},& j = k+1.
\end{cases}
\end{equation}
In order to proceed, we bring the implicit part of the update over to the RHS and assume that $\epsilon$ is sufficiently small to expand the inverse matrix. The relevant quantitative criterion here is that the magnitude of the eigenvalues of $\epsilon\theta M$ are smaller than 1, i.e., the $\max\left[ |\lambda_1|,|\lambda_2|,... \right] < 1$. In turn, we obtain
\begin{align}
 \phi^{\lambda+1} &=  \left( I - i \epsilon\theta M \right)^{-1}\left\{ (I + i \epsilon(1-\theta)M)\phi^{\lambda} + \sqrt{\epsilon}\eta^\lambda \right\}\\
& = \sum_{k=0}^\infty \left( i \epsilon \theta M \right)^k \left\{ (I + i \epsilon(1-\theta)M)\phi^{\lambda} + \sqrt{\epsilon}\eta^\lambda \right\}.
\end{align}
Let us truncate the expansion at second order in $\epsilon$ and focus on the contributions to the drift term
\begin{align}
    \phi^{\lambda+1} = \left\{ \Big(1 + i \epsilon M - \epsilon^2\theta M^2 \Big)\phi^{\lambda}  + \sqrt{\epsilon}\eta^\lambda \right\} +{\cal O}(\epsilon^{3/2}). \label{eq:dereffcompl}
\end{align}
The correction to the drift term of second order in $\epsilon$ may be absorbed into an effective action for the general EM scheme
\begin{align}
S_{\theta} = \frac{1}{2}\phi \Big( M +  i\epsilon \theta M^2 \Big) \phi = S_{\textrm{explicit}} + \frac{i\epsilon}{2} \theta \sum_j S_j^2. \label{eq:effaccompl}
\end{align}
The expression for $S_\theta$ tells us that the difference between the EM scheme with finite $\theta$ and the fully explicit one lies in the presence of one additional term. It is proportional to the complex unit $i$ and both depend on the time step and the implicitness parameter. Similar to the regulator term from a tilted contour in \cref{eq:tiltreg} it is positive and thus leads to a damping of the oscillations of the path integral. \cref{eq:effaccompl} thus constitutes a new means of regularization unavailable to explicit solvers.

The above argument is further supported by numerical tests, which show that the regularization becomes weaker as the Langevin time step is reduced. Intuitively it also agrees with the behavior of the numerical solvers we discussed in the context of large excursions in \cref{sec:excusions}. The term proportional to $\epsilon^2$ in \cref{eq:dereffcompl} features a minus sign, which leads to the stable undershooting of the true solution shown in  \cref{fig:FlowPhi4Mod}. In turn, it is this correction that will prevent the Langevin dynamics from diverging in the late Langevin time limit, realizing the role of a regularizer in the underlying path integral.

\subsection{Finite Langevin time step errors}

We have argued in the preceding section that implicit solvers provide a novel intrinsic regularization of the underlying complex path integral. This regularization depends on the implicitness parameter but more importantly depends on the finite Langevin time step $\Delta \lt$. It tells us that for finite step size our system remains well defined but that when moving towards continuous Langevin time, the dynamics will become more difficult to tame, as we concurrently remove our regularization. This conundrum can be avoided if it is possible to analytically correct for the finite Langevin step size corrections in our observables. Then we may choose a small but not too small value of $\Delta \lt$ (depending on the parameters of the system) and carry out the simulation in a well-defined manner, accounting for the difference to the continuous Langevin solution a posteriori. In this section, we set out to derive such correction terms.

Our strategy is as follows: As a first step, we follow \cite{Kronfeld:1992jf} and show that the effects of an implicit solver scheme at finite Langevin step size $\Delta \lt$ can be cast in the language of an effective action for the Fokker-Planck equation. In order to exploit the well-established methods underlying the derivation of the Fokker-Planck equation from Langevin dynamics, we restrict ourselves to a scenario with purely real Feynman weights, i.e. the imaginary time one. We continue in a second step to guess how the effective action obtained in the real case generalizes to the complex case. This heuristic step is supported by numerical evidence, which confirms that it allows us to correct numerical artifacts introduced by finite $\Delta \lt$ in practice.

Let us again focus on the simple but relevant free theory with $V(\phi) = \frac12 m^2\phi^2$. The update step of the general EM scheme, now for the Euclidean action reads
\begin{equation}\label{eq:EMSchemeSQMEucl}
    \phi_j^{\lambda + 1} = \phi_j^{\lambda} - \epsilon_j\left[ \theta \frac{\partial S^{\lambda +1}}{\partial \phi_j} + (1-\theta)\frac{\partial S^{\lambda}}{\partial \phi_j}  \right] + \sqrt{\epsilon_j}\eta_j^\lambda,
\end{equation}
where $\epsilon_j=\frac{\Delta \lt}{|\omega_j|}$ and the negative sign in the drift term arises from the Wick rotation into imaginary time. For $\theta=0$ these dynamics have been investigated in ref.\cite{Kronfeld:1992jf}.

To simplify the derivation below, we assume without loss of generality that the step size $\epsilon_i = \epsilon$ is constant along the contour. Remember that in the free theory can write the action in a simple matrix form as $\frac{\partial S^{\lambda}}{\partial \phi_i} = M\phi^\lambda$. Substituting all into \cref{eq:EMSchemeSQMEucl} yields
\begin{equation}
    \left( I + \epsilon\theta M \right)\phi^{\lambda+1} =  \left\{ (I - \epsilon(1-\theta)M)\phi^{\lambda} + \sqrt{\epsilon}\eta^\lambda \right\}.
\end{equation}
Let us bring the implicit part of the update over to the RHS and take $\epsilon$ is small enough to expand the first term in parentheses
\begin{align}
 \phi^{\lambda+1} &=  \left( I + \epsilon\theta M \right)^{-1}\left\{ (I - \epsilon(1-\theta)M)\phi^{\lambda} + \sqrt{\epsilon}\eta^\lambda \right\}\\
& = \sum_{k=0}^\infty \left( -\epsilon \theta M \right)^k \left\{ (I - \epsilon(1-\theta)M)\phi^{\lambda} + \sqrt{\epsilon}\eta^\lambda \right\}.
\end{align}
The above expression, up to order $\epsilon^{5/2}$, can be written in index notation, using $M_{jk}\phi_k^\lambda = \frac{\partial S^\lambda}{\partial \phi_j} = S_j^\lambda$ as
\begin{equation}
    \phi^{\lambda + 1}_j = \phi^{\lambda}_j - \epsilon S^\lambda_j + \epsilon^2\theta M_{jk} S^\lambda_k + \left(\sqrt{\epsilon}\delta_{jk} - \epsilon^{3/2}\theta M_{jk} \right) \eta_k^\lambda + \mathcal{O}(\epsilon^{5/2}) = \phi_j^\lambda - f_j^\lambda[\phi].
\end{equation}

We will now derive the corresponding Fokker-Plank equation and the effective action based on the above update prescription. The standard approach (see e.g. \cite{Zinn-Justin:572813}) is to rewrite the probability distribution for $\phi$, denoted as ${\cal P}[\phi]$ at discrete Langevin time step $\lambda+1$ in terms of its values at step $\lambda$ using a delta-distribution. The argument of the delta distribution contains the Langevin update step from $\lambda$ to $\lambda+1$ and is averaged over the ensemble
\begin{equation}
\mathcal{P}^{\lambda+1}[\phi] = \int [d\phi'] \left\langle \prod_j \delta \left(\phi_j - \phi_j' + f_j[\phi']\right) \right\rangle \mathcal{P}^\lambda[\phi'].
\end{equation}

After expanding the delta function in powers of $f_j$ and integrating over $\phi'$ one arrives at the Kramers-Moyal expansion for the discretized stochastic process, 
\begin{equation}
    \mathcal{P}^{\lambda+1}[\phi] = \mathcal{P}^{\lambda}[\phi] + \sum_{n=1}^\infty \frac{1}{n!}\nabla_{j_1}...\nabla_{j_n} \left( \langle f_{j_1} ... f_{j_n} \rangle \mathcal{P}^\lambda[\phi] \right).
\end{equation}
A Fokker-Planck equation may be obtained by considering terms up to the order $\epsilon^2$, which are encoded in the correlation functions of the update term $f$.
To make these explicit we use the following properties of the noise $\langle \eta_j \rangle = 0$, $\langle \eta_j \eta_k \rangle = 2\delta_{jk}$, $\langle \eta_j \eta_k \eta_l \rangle = 0$ and $\langle \eta_j \eta_k \eta_l \eta_m \rangle = 4\left( \delta_{jk}\delta_{lm} + \delta_{jl}\delta_{km} + \delta_{jm}\delta_{kl} \right)$, which leads to the following four expressions: 
\begin{equation}
\begin{aligned}
\langle f_j \rangle  = & \epsilon S_j - \epsilon^2\theta M_{jk}S_k + \mathcal{O}(\epsilon^3), \\
\langle f_j f_k \rangle =& \epsilon^2 S_j S_k + 2\epsilon \delta_{jk} - 2\epsilon^2\theta(M_{kl}\delta_{jl} +  M_{jl}\delta_{kl}) + \mathcal{O}(\epsilon^{5/2}), \\
\langle f_j f_k f_l \rangle = &  2\epsilon^2 \left( S_j\delta_{kl} + S_k \delta_{jl} +  S_l \delta_{jk} \right), \\
\langle f_j f_k f_l f_m \rangle =&4\epsilon^2 \left( \delta_{jk}\delta_{lm} + \delta_{jl}\delta_{km} + \delta_{jm}\delta_{kl} \right).
\end{aligned}\label{eq:corrsoff}
\end{equation}
To the lowest order in $\epsilon$ we obtain the following Fokker-Planck equation
\begin{equation}
    \frac{\partial}{\partial \lt}\mathcal{P} = \nabla_j \left[\left( S_j + \nabla_j \right) \mathcal{P}\right] + {\cal O}(\epsilon^{3/2}), \label{eq:FPdiscre1}
\end{equation}
which by a change of variable, $\mathcal{P}=e^{-S/2}\Psi$, can be shown to converge to the correct equilibrium distribution in the case of a real-valued action. Now the Kramer-Moyal expansion up to corrections of order ${\cal O}(\epsilon^3)$ on the other hand contributes additional terms to the Langevin time evolution of the probability distribution
\begin{equation}
\begin{aligned}
\partial_t \mathcal{P} = \nabla_j \left( S_j + \nabla_j \right) \mathcal{P} + \epsilon\left\{-\theta M_{jk}\nabla_j(S_k \mathcal{P}) + \frac12\nabla_j\nabla_k (S_jS_k \mathcal{P}) \right. \\ 
\left. - \theta (M_{kj} + M_{jk}) \nabla_j\nabla_k\mathcal{P} + \nabla_j\nabla^2 (S_j\mathcal P) + \frac12 \nabla^2 \nabla^2 \mathcal{P} \right\} + {\cal O}(\epsilon^{5/2}).\label{eq:FPdiscre2}
\end{aligned}
\end{equation}
Since it is the equilibrium distribution, which is of main interest to us, let us set the LHS to zero. To be more concise we will rewrite $M_{jk} = -\nabla_j S_k$ and use \cref{eq:FPdiscre1} to make the replacement $\nabla_j \mathcal{P} = -S_j\mathcal{P} + \mathcal{O}(\epsilon)$ within the curly brackets of \cref{eq:FPdiscre2}, consistent with the order of the approximation. The new terms at this order then give
\begin{align}
&\nabla_j\left\{\theta M_{jk}S_k  + \frac12 ( S_{jk}S_k + S_jS_{kk} - S_j S_k^2) - \theta(M_{kj} + M_{jk})S_k \right. \\ 
& \left. + \nabla_k(S_{jk}-S_jS_k) + \frac12\nabla_j(-S_{kk} + S_k^2) \right\}\mathcal{P} \\
=&\nabla_j\left\{ -\theta M_{kj}S_k  + \frac12 ( S_{kj}S_k + S_jS_{kk} - S_j S_k^2) \right. \\
& \left. + (S_{jkk} - S_{jk}S_{k} - S_{jk}S_k - S_jS_{kk} + S_jS_k^2) \right. \\
& \left. + \frac12(-S_{kkj} + S_{kk}S_j + 2S_{kj}S_k - S_k^2 S_j) \right\}\mathcal{P} \\
=&\nabla_j\left\{ \theta S_{jk}S_k  -\frac12 S_{jk}S_k  + \frac12 S_{jkk}
\right\}\mathcal{P} = \nabla_i\left\{ -\left( \frac12 - \theta \right) S_{jk}S_k  + \frac12 S_{jkk}
\right\}\mathcal{P} \\ 
=& \nabla_j\left\{\frac12  \nabla_j  S_{kk} - \frac12 \left( \frac12 - \theta \right) \nabla_j S_k^2 \right\}\mathcal{P}. \label{eq:FPdiscre2final}
\end{align}
Reexpressed as a modified action we arrive at the intermediate result
\begin{equation}
0=\nabla_j \left[\left( \bar S_j + \nabla_j \right) \mathcal{P} \right], \quad
    \bar{S} = S + \frac{\epsilon}{2}\sum_k\left\{S_{kk} -\left( \frac12  - \theta\right) S_k^2  \right\}.\label{eq:FPdiscrreal}
\end{equation}
In the above expression, we see that the $\theta$ parameter governs the size and sign of a real-valued addition to the action. For $\theta>\frac12$ the contribution is positive and for $\theta<\frac12$ it is negative, distinguishing clearly between the implicit regime and the explicit regime. Note that similar to our discussion of the large excursions, the semi-implicit case of $\theta=\frac12$ is special, as it cancels all corrections associated with the $S_k^2$ term.

\begin{figure}[t]
\centering 
        \includegraphics[scale=0.40]{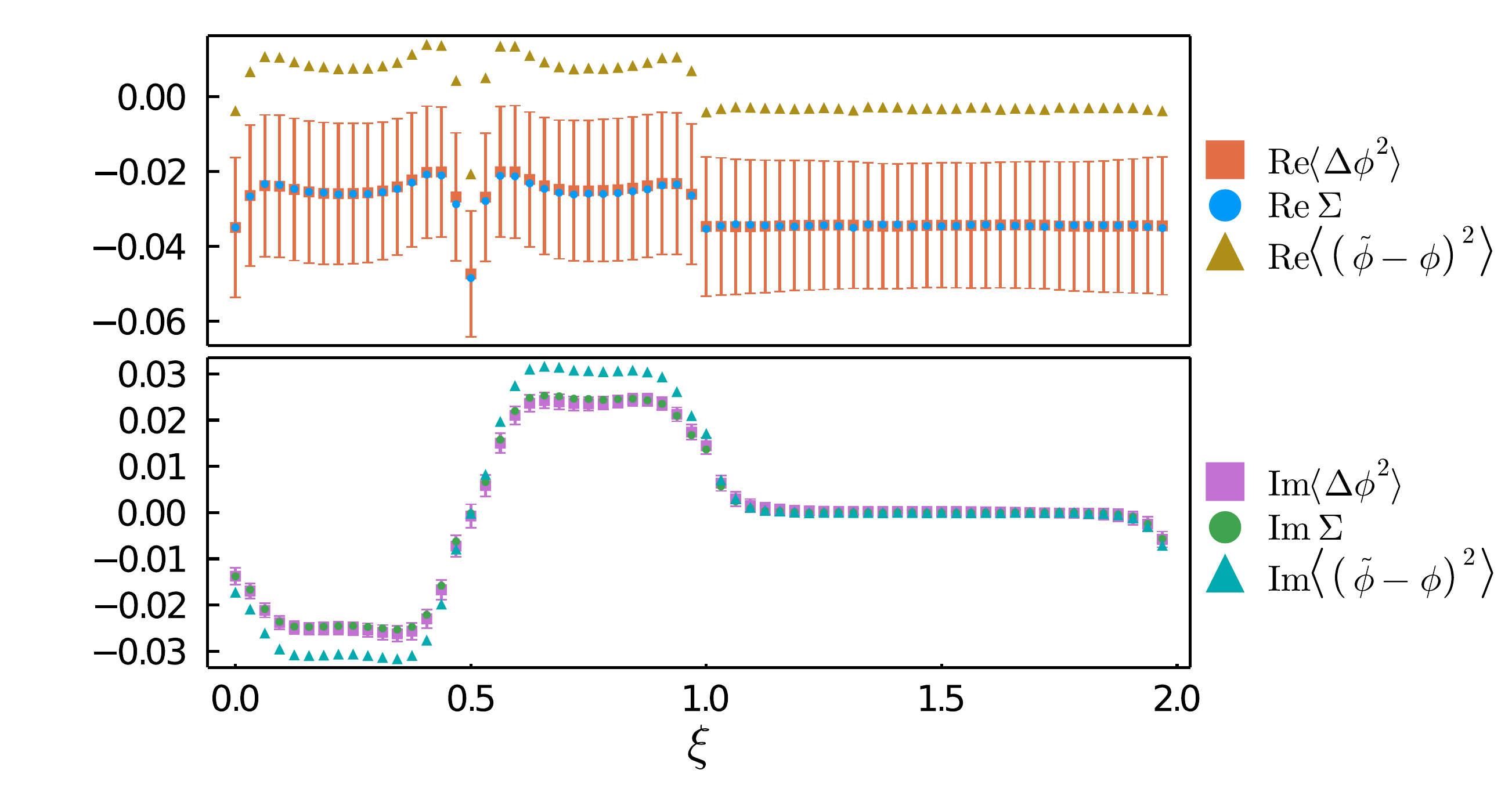}
        \includegraphics[scale=0.42]{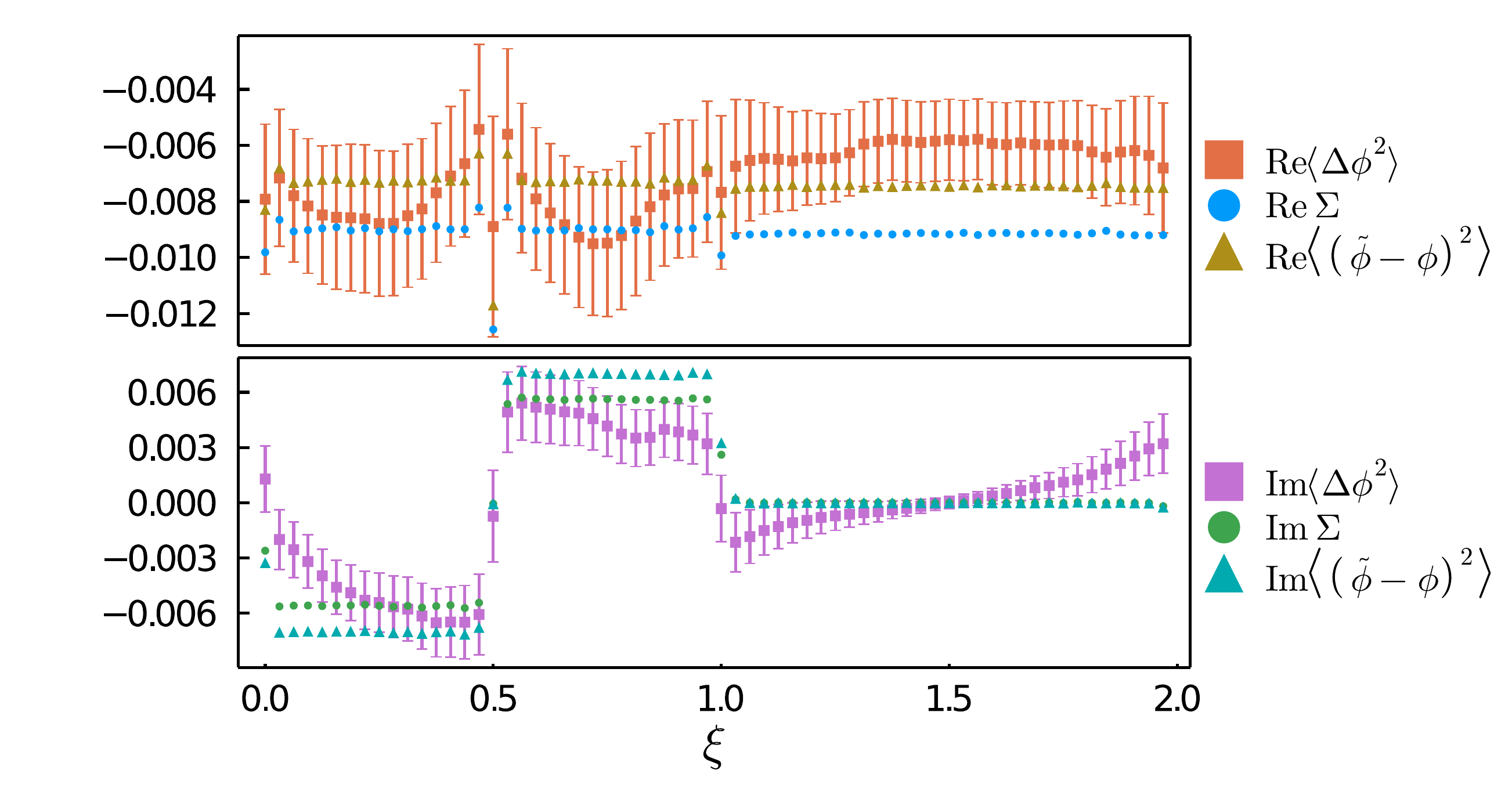}
    \caption{ Comparison of the finite Langevin step-size artifacts in the equal-time correlator $\Delta \phi^2=\langle \phi^2\rangle_{\rm CL} - \langle \phi^2 \rangle_{\rm QM}$ to the estimates of that error based on \cref{eq:corrphi2}. The filled triangles denote the full estimate including the terms proportional to $\epsilon^2$, which provide the correct qualitative behavior but systematically overestimate $\Delta \phi^2$. On the other hand the leading order expression $\Sigma$, proportional to $\epsilon$, shown as filled circles captures the error even quantitatively within the statistical uncertainty. The position along the contour is parametrized by $\xi$, which for $\xi<1$ points to real-time values and for $1<\xi<2$ refers to imaginary times. (top) Estimation of the errors in the free theory (harmonic oscillator) using $\Delta \lt=10^{-2}$, as well as (bottom) for the anharmonic oscillator at $\lambda = 24$ with $\Delta \lt=10^{-3}$. In both cases, the data is based on 1000 separate trajectories each of total length $\lt=200$.}
    \label{fig:modActionCorrectionHO}
\end{figure}

The derivation outlined above cannot be translated one-to-one into the complex case. We would need to instead express the complex Langevin evolution in terms of the real-valued joint probability distribution of the real- and imaginary part of the complexified fields. Doing so, we were unable to derive a similarly closed-form as \cref{eq:FPdiscre2final}. The structure of the correction terms obtained in the real case however invites a heuristic generalization to the complex domain using the replacement $-S\to iS$, which leads to
\begin{equation}
0=\nabla_j \left[\left( -i\bar S_j + \nabla_j \right) \mathcal{P} \right], \quad
    \bar{S} = S + \frac{\epsilon}{2}\sum_k\left\{S_{kk} + i\left( \frac12  - \theta\right) S_k^2  \right\}. \label{eq:FPdiscrcmplx}
\end{equation}
Let us find out whether this expression describes the dynamics of the complex Langevin simulation in practice. Similar to the discussion for the real-valued case in \cite{Kronfeld:1992jf}, we can attempt to counteract the effects introduced by a finite Langevin step size $\epsilon$ in the action by a redefinition of the fields. In our case the leading order change in fields according to \cref{eq:FPdiscrcmplx} amounts to  
\begin{equation}\label{eq:corrected_field}
\tilde  \phi_j = \phi_j - \frac{i\epsilon}{2}\left( \frac12 - \theta \right)S_j,
\end{equation}
where $S_j$ is nothing but the drift term. Note that $S_{jj}$ in the free theory is just a constant, so that acting with one more derivative on it makes that term vanish. Thus the redefinition of the fields changes the action to first order in $\epsilon$ such that it cancels the $S_j^2$ contribution in \cref{eq:FPdiscrcmplx}
\begin{equation}
    \bar S[\tilde \phi] \sim \bar S[\phi]  - \frac{i\epsilon}{2}\left(\frac12 - \theta\right)\sum_k S_k^2.
\end{equation}
I.e. if \cref{eq:FPdiscrcmplx} is the correct generalization then observables evaluated in terms of $\tilde\phi_i$ instead of $\phi$ should show reduced deviations from the continuous Langevin time result. For the equal time two-point function we e.g. obtain the following corrected expression
\begin{equation}
    \left\langle \tilde\phi_j^2 \right\rangle  = \left\langle \phi_j^2 \right\rangle \underbracket{-   \epsilon\left( \frac12 - \theta \right)\left\langle \phi_j (iS_j) \right\rangle}_{\Sigma} - \frac{\epsilon^2}{4}\left( \frac12 - \theta \right)^2 \left\langle S_j^2 \right\rangle. \label{eq:corrphi2}
\end{equation}
For later reference, we denote the correction term linear in $\epsilon$ as $\Sigma$. In order to assess the validity of the above arguments, let us simulate the harmonic oscillator on the real-time contour c) of \cref{fig:contours}, i.e. on a contour without tilt in the real-time branch, up to a maximum extent of $x_0^{\rm max}=0.5$. Deploying an equidistant real-time spacing on the forward and backward branch of the contour and a Langevin time step of $\Delta \lt=10^{-2}$, we compute the difference between the numerical result and the analytic solution $\Delta \phi^2=\langle \phi^2\rangle_{\rm CL} - \langle \phi^2 \rangle_{\rm QM}$ as the colored boxes in the top panel of \cref{fig:modActionCorrectionHO} vs. the contour parameter $\xi$. We find characteristic features in both the real- and imaginary part of this quantity. The artifacts introduced by the implicit solver at finite Langevin time show opposite sign in the imaginary part and same sign in the real-part comparing the forward and backward branch. On the Euclidean time interval, only the real-part receives significant modifications.

Interestingly, the correction term $(\phi-\tilde\phi)^2$ taken from \cref{eq:corrphi2}, when plotted as the colored triangles in \cref{fig:contours} already follows the behavior of the deviations in a qualitative fashion. At the same time, we observe that it appears to consistently over-predict those artifacts. Limiting ourselves to the corrections linear in Langevin step size $\epsilon$, shown in gray, we find that they capture the artifacts even more accurately. This difference between the linear and quadratic terms in $\epsilon$ to us hints at the need to include higher-order corrections in the expansion of \cref{eq:FPdiscrcmplx} to arrive at a reliable correction term beyond leading order. To conclude, we find that the linear correction terms derived from a heuristic generalization of the robust result in \cref{eq:FPdiscrreal} capture the discrete dynamics of our complex Langevin simulation in a qualitative fashion, lending numerical support to \cref{eq:FPdiscrcmplx}.

The correction terms obtained in \cref{eq:FPdiscrcmplx} contain the first and second derivative of the action $S_k$ and $S_{kk}$. We can gain additional insight into what role they play from the following considerations based on the translation invariance of the integrals over the fields. Let us start by stating the fact that 
\begin{align}
   \nonumber  \langle (\phi _j +\phi_0)^n \rangle =& \int D \phi (\phi _j +\phi_0)^n \exp(iS(\phi_j)) \\
     =& \int D \phi (\phi _j)^n \exp(iS(\phi_j-\phi_0))
\end{align}
where we have shifted the integral in the second line, such that the constant $\phi_0$ has been moved into the exponent. While the distribution obtained from CL is different, the expectation values of the shifted system will still be the same. 
We can now exploit the presence of $\phi_0$ in the integrand and the weight to derive relations between different n-point correlation functions. Using the first and second derivative with respect to $\phi_0$ we have
\begin{align}
\nabla_{\phi_{0}}\langle (\phi _j +\phi_0)^n \rangle |_{\phi_0 =0}  =& n \langle (\phi _j)^{n-1} \rangle =  \langle (\phi _j )^n (-iS_j) \rangle \\
\nabla_{\phi_{0}} ^2 \langle (\phi _j +\phi_0)^n \rangle |_{\phi_0 =0}  =& n (n-1) \langle (\phi _j)^{n-2} \rangle =  \langle (\phi _j )^n (-iS_{jj} -(S_i)^2) \rangle.
\end{align}
Using the first derivative with $n=1$ and the second derivative with $n=0$ we obtain the following two expressions respectively
\begin{equation}
    \langle 1 \rangle =  \langle \phi _j  (-iS_j) \rangle, \qquad 0 =  \langle  -iS_{jj} -(S_i)^2 \rangle.
\end{equation}
For the free case, $S_{jj}$ is a constant. The first term is particularly interesting as it tells us that for continuous Langevin time, the term  $\langle \phi _j  (-iS_j) \rangle$ should be constant and correspond to the normalization of the system. In the presence of discrete Langevin time steps, we found that it is a term proportional to $\langle \phi _j  (-iS_j) \rangle$, which describes the corrections and which are not constant as shown in \cref{fig:modActionCorrectionHO}. We thus interpret the corrections in \cref{eq:corrphi2} as counteracting in part the deviations from the correct normalization of the continuum theory.

When we derived the modifications to the Fokker-Planck equation in the free theory, we were able to express them in terms of the quantity $S_j$. We may ask whether this expression also holds in the interacting theory. To this end, we carry out simulations of the anharmonic oscillator at short real-times on the untilted Schwinger-Keldysh contour up to $x_0^{\rm max}=0.5$, where complex Langevin is known to converge to the correct solution (for more details see \cref{sec:numsimtherm}). The same solver as for the harmonic oscillator is deployed and we choose a Langevin step-size of $\Delta\lt=10^{-3}$. In the lower panel of \cref{fig:modActionCorrectionHO} we plot the resulting deviations from the analytic solution (filled boxes), compared to the naive application of \cref{eq:corrphi2} to the interacting theory (filled triangles). Again we observe that the expression up to second order in $\epsilon$ slightly overestimates the artifacts but that restricting us to the linear term in $\epsilon$ allows us to capture the discretization errors within uncertainties.

With an expression at hand that allows us to correct the finite Langevin step size corrections for small values of $\epsilon$, we are able to exploit the regularization properties of the implicit solvers in practice. What remains for each explicit system is to choose a step-size and $\theta$ parameter, keeping in mind the trade-off between regularization artifacts and numerical cost. Having too small of a step size in the implicit scheme will reduce the effect of the regulator, which in turn will lead to the appearance of large excursions. Even though these excursions do not represent a problem in principle (the approach is inherently stable) they may lead to high computational cost if a fixed accuracy goal is prescribed. Using an intermediate step sizes $\sim10^{-3}$ appears to give the best trade-off for the interacting systems considered in this study. The finite step size provides an effective regulator to the path integral and the finite step-size artifacts can be remedied by the correct procedure discussed above.

We emphasize that the implicit EM scheme provides enough of an intrinsic regularization that we may forego a tilting of the Schwinger-Keldysh contour all together. I.e. we gain access to the fields very close to the actual forward and backward real-time branch of the canonical Schwinger-Keldysh contour, which is particularly useful in the study of non-equilibrium field theory, in which the forward and backward correlators are not related via the KMS relation. 

Armed with the insight laid out in the previous sections we are now ready to carry out stable simulations of the quantum anharmonic oscillator at short real times.

\section{Stable CL simulations at short real-times}
\label{sec:numsim}
In this section, we present numerical results of simulating real-time complex Langevin on the canonical Schwinger-Keldysh contour with short time extent of $x_0^{\rm max} =0.5$ using the implicit EM scheme.
We will start out with a system in thermal equilibrium which is formulated on contour c) of \cref{fig:contours}. As a second example, we take a look at a system with Gaussian initial conditions, where only the forward and backward real-time branch of the contour remains and a Euclidean branch is absent.

\subsection{Dynamics in thermal equilibrium}
\label{sec:numsimtherm}

Our simulation uses the same parameters as adopted in the classic work of ref.\cite{BergesSexty2007}, i.e. $\lambda=24$ and $m=1$. To discretize the real-time contour c) in \cref{fig:contours} for a temperature $T=1/\beta=1$ and real-time extent of $x_0^{\rm max} =0.5$, we use 16 points for the forward and backward branch each and an additional 32 points along the negative imaginary time axis. This choice of an equidistant $|a|=0.031$ guarantees that the finite time spacing artifacts to the correlation functions remain at the permille level. 

To regularize the path integral we deploy the general EM scheme with its implicitness parameter set to $\theta=0.6$. We use the adaptive step size prescription of the Julia stochastic processes library with a maximum step size of $\Delta \lt=0.005$. This choice provides an efficient enough regularization to avoid costly excursions, while at the same time the deviation from the continuous Langevin result remains smaller than our statistical uncertainty. For the computation of the correlation functions of interest, field configurations are collected based on 500 different trajectories. We read out observables on each of them in intervals of $\delta \lt =0.1$ up to a total Langevin time of $\lt =100$.
\begin{figure}\centering 
        \includegraphics[scale=0.45]{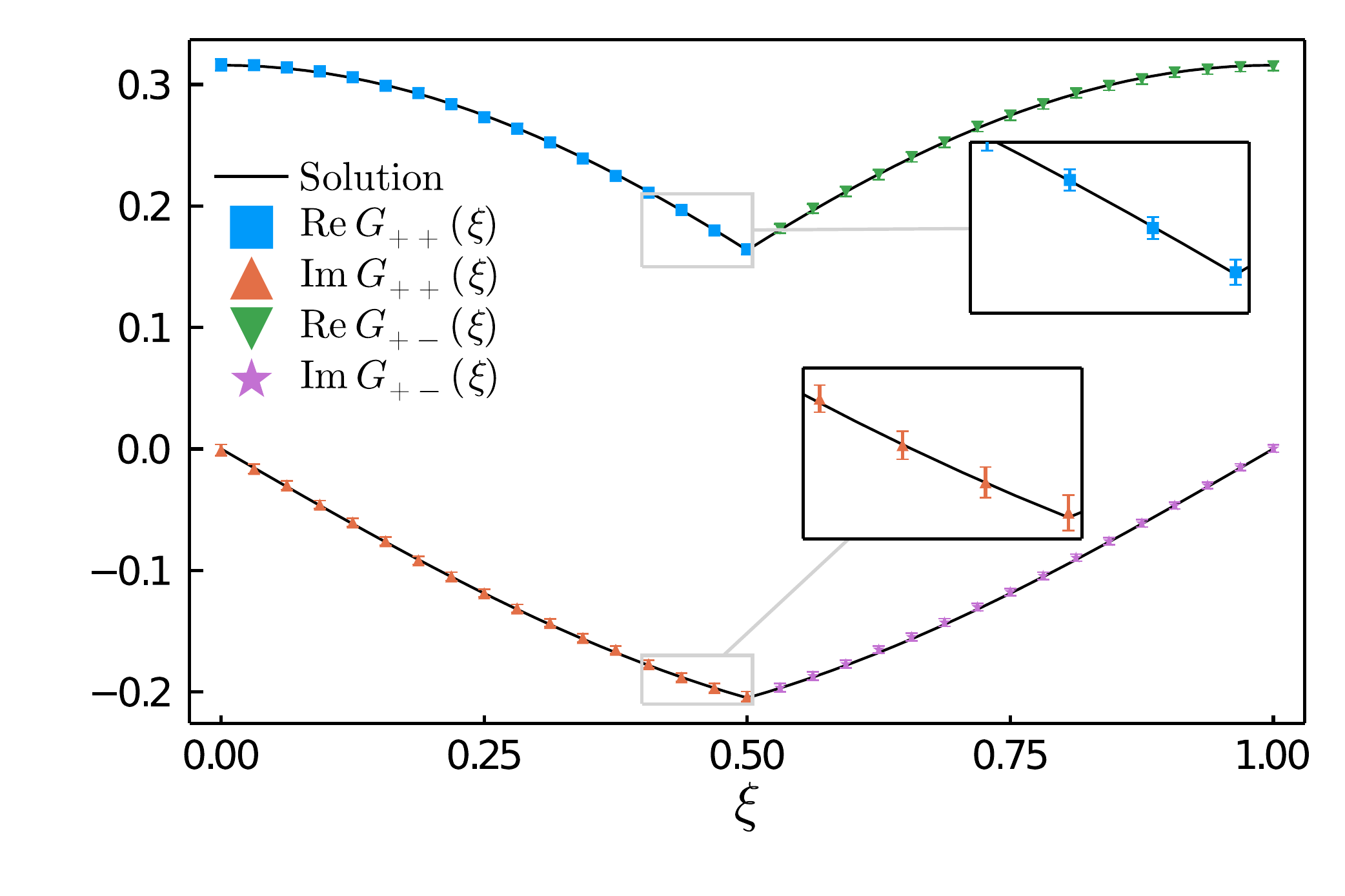}
        \includegraphics[scale=0.45]{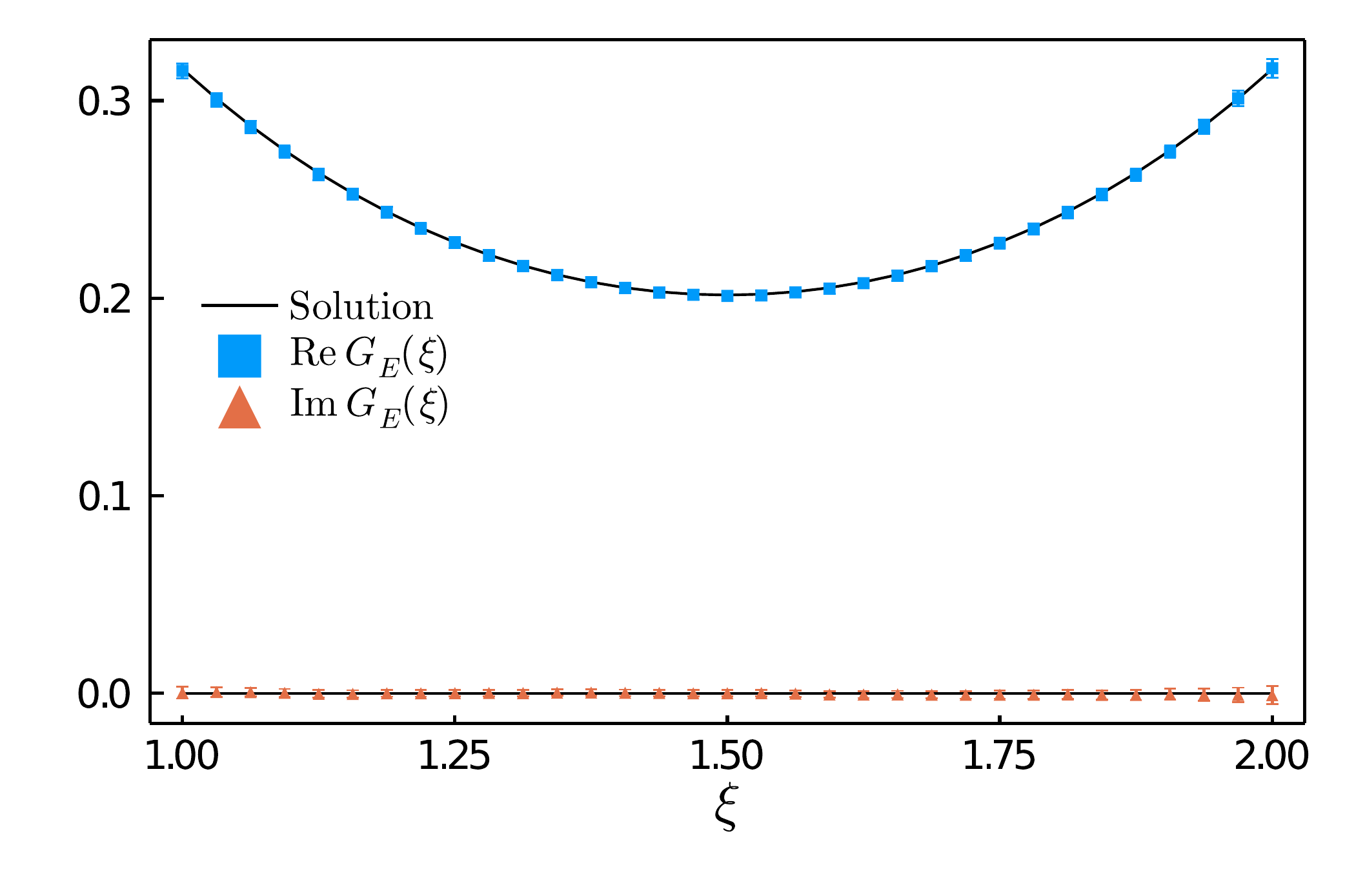}
    \caption{(top) Two unequal time correlation functions along the real-time branches of the canonical Schwinger-Keldysh contour. For $\xi<1/2$ the datapoints represent $G_{++}(x_0=\xi)=\langle\phi(0)\phi(\xi)\rangle- \langle \phi(0)\rangle\langle\phi(\xi)\rangle$ and for $\xi>1/2$ we have $G_{+-}(x_0=1-\xi)=\langle\phi(0)\phi(\xi)\rangle- \langle \phi(0)\rangle\langle\phi(\xi)\rangle$. Note the excellent accuracy in reproducing the continuous Langevin time result given as black solid line. (bottom) The Euclidean correlator $G_E(x_0=-i(\xi-1))=\langle\phi(0)\phi(\xi)\rangle$ evaluated on the imaginary time branch of the canonical Schwinger-Keldysh contour together with the continuous Langevin time solution (black solid)}\label{fig:conoical_SK_therm_twoP_func}
\end{figure}

In the top panel of \cref{fig:conoical_SK_therm_twoP_func} we show the unequal time correlation function $G=\langle \phi(0)\phi(\xi)\rangle - \langle \phi(0)\rangle\langle\phi(\xi)\rangle$, which for $\xi<1/2$ amounts to $G_{++}(x_0=\xi)=\langle\phi(0)\phi(\xi)\rangle - \langle\phi(0)\rangle\langle\phi(\xi)\rangle$ and for $\xi>1/2$ to $G_{+-}(x_0=1-\xi)=\langle\phi(0)\phi(\xi)\rangle-\langle\phi(0)\rangle\langle\phi(\xi)\rangle$. The corresponding continuum Langevin time solution is plotted as solid black curve, which is obtained from a matrix mechanics computation based on the truncated Hilbert space spanned by the lowest 32 energy eigenstates of the harmonic oscillator. The magnified insets confirm that our solution accurately reproduces the continuum solution on the forward and backward branch of the canonical Schwinger-Keldysch contour up to these early real-times.

In the lower panel of \cref{fig:conoical_SK_therm_twoP_func} we plot the Euclidean correlator $G_E(x_0=-i(\xi-1))=\langle\phi(0)\phi(\xi)\rangle$. It features a vanishing imaginary part and a real part which correctly exhibits a symmetry around $x_0=-i\beta/2$, corresponding to the contour parameter $\xi=1.5$ here. Again the continuum solution from matrix mechanics is given as solid black line and we find excellent agreement.
\begin{figure}
    \centering
    \includegraphics[scale=0.4]{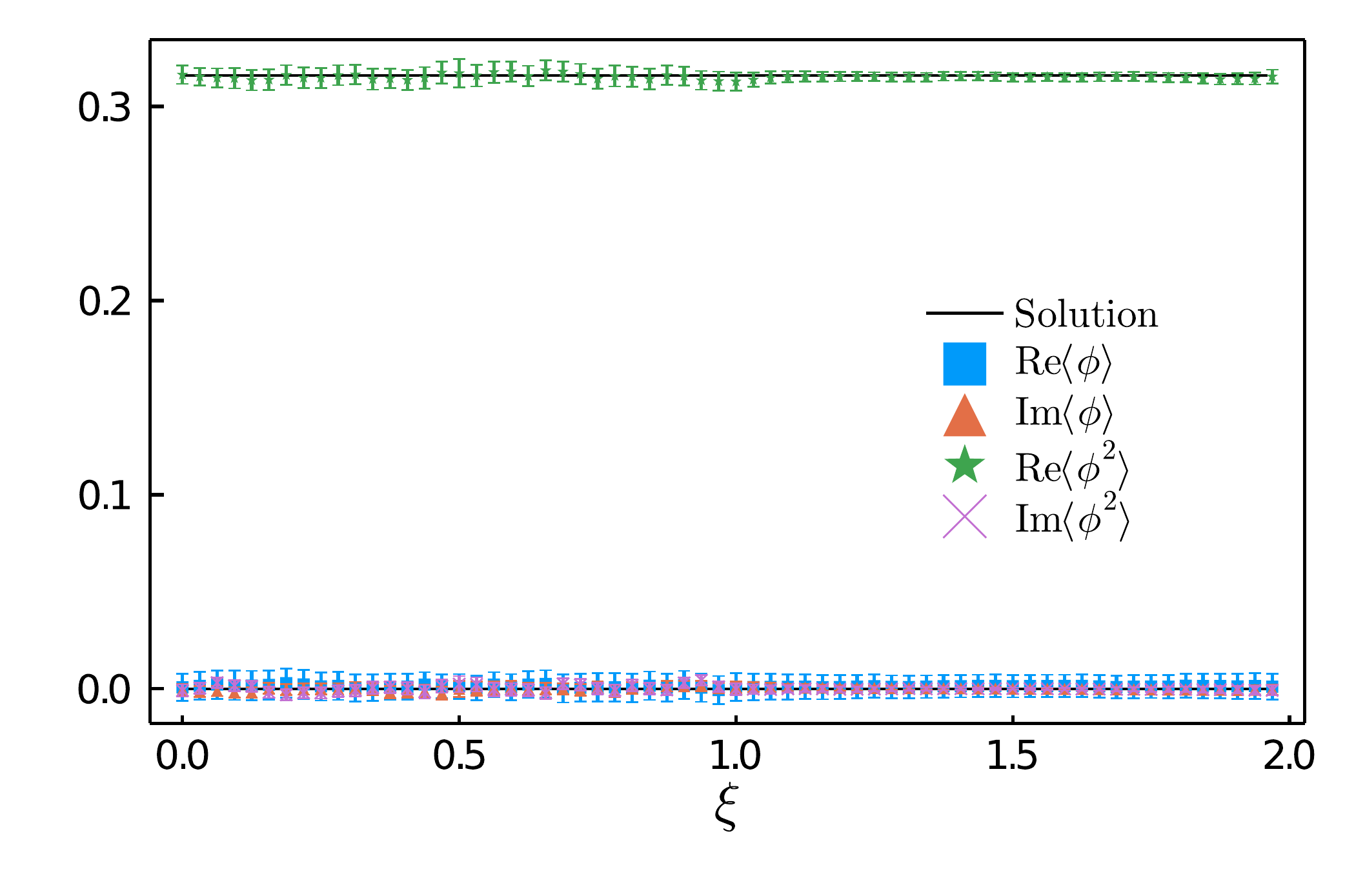}
    \includegraphics[scale=0.4]{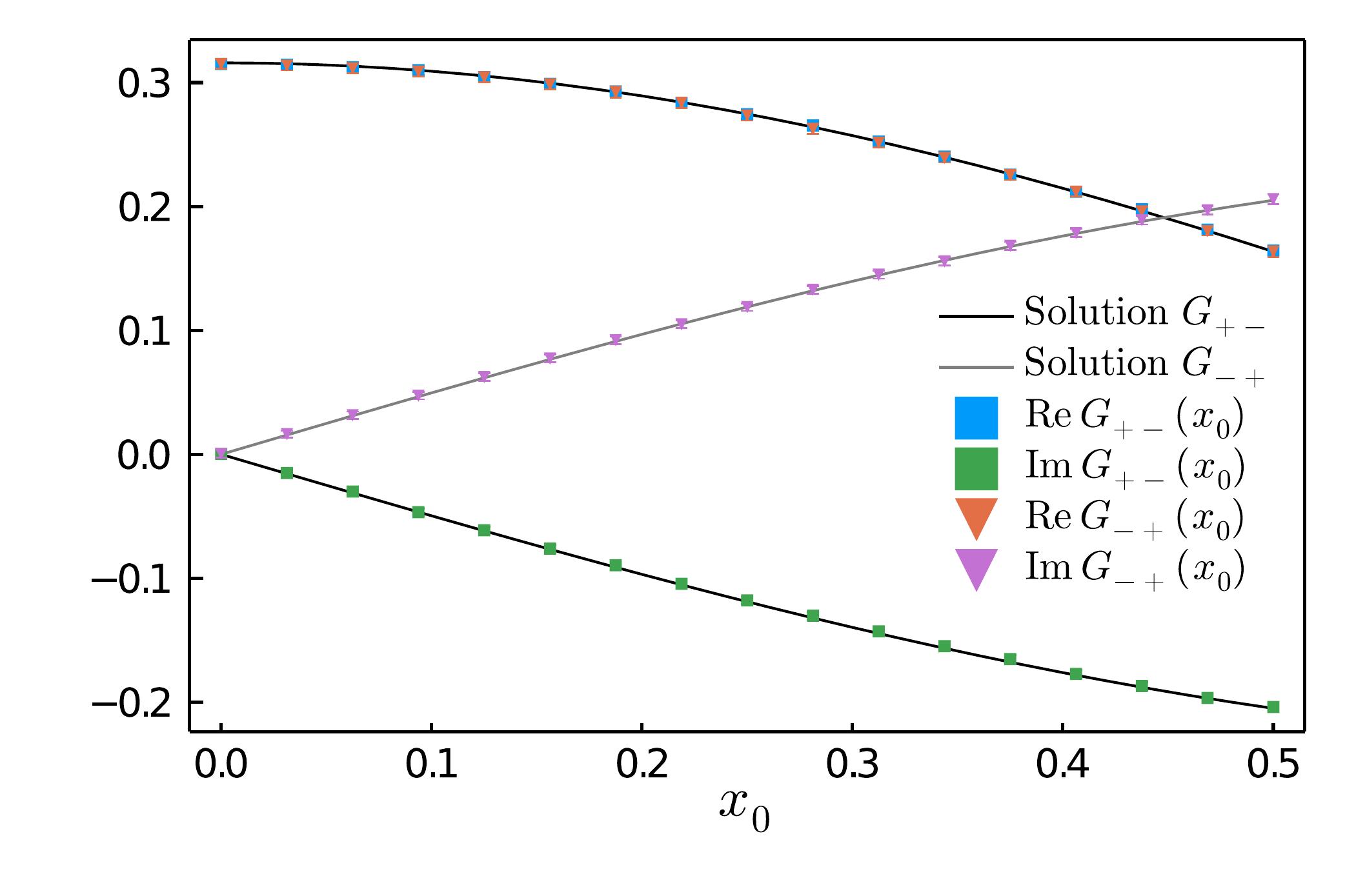}
    \caption{(top) The field expectation value $\langle \phi\rangle$ (box and triangle), as well as the equal-time correlation function $\langle \phi^2\rangle$ (filled star and cross) evaluated in thermal equilibrium along the whole simulation contour, parametrized by $\xi$. The continuous Langevin time solution from matrix mechanics is given as solid line. (bottom) The phenomenologically relevant forward $G_{+-}=G^>$ and backward $G_{-+}=G^<$ correlation functions. Note that only in the thermal setting the information they contain is redundant with that of the $G_{++}$ correlator.}
    \label{fig:cononical_SK_therm_oneP_func}
\end{figure}

Let us take a look at another set of observables, which have been discussed in the literature. The top panel of \cref{fig:cononical_SK_therm_oneP_func} we show the field expectation value $\langle \phi\rangle$ (filled box and triangle) and the equal time correlation function $\langle \phi^2\rangle$ (filled star and cross) along the whole extent of the simulation contour parametrized by $\xi$. Within the statistical uncertainties of our simulation, we find full agreement with the continuous-time Langevin solution. Should one be interested in higher precision results, one will eventually find minute differences from the continuum result, similar to those shown in the lower panel of \cref{fig:modActionCorrectionHO}. 

We foresee that the new insight obtained in \cref{eq:corrphi2} will help us in future studies to distinguish artifacts arising from finite Langevin-time discretization from those connected to a convergence to the wrong result. One concrete example is the equal-time correlation function, whose deviation from a constant value has previously been taken as an indication for the arrival at an unphysical solution. If the errors from a finite $\Delta \lt$ are accounted for, can the remaining deviation be unambiguously associated with wrong convergence.

The last result in this section we present in the lower panel of \cref{fig:cononical_SK_therm_oneP_func}. In preparation for the simulation of genuine field theory in higher dimensions and for the simulation out-of-equilibrium in the next section, we compute the forward $G_{+-}(x_0)=G^>(x_0)$ and backward correlator $G_{-+}(x_0)=G^<(x_0)$ together with the analytic solution as solid lines. In thermal equilibrium, the information of these two quantities is redundant due to the KMS relation and because we are in a quantum mechanical setting their interrelation is actually trivial. Their real parts agree, while their imaginary parts are the negative of each other. They nevertheless take on a central role in field theory, as their difference $\rho=G^>-G^<$ encodes the spectral function of the system, which harbors a wealth of phenomenologically relevant pieces of information. Thus an accurate reproduction of these correlation functions between fields on different branches of the contour is an important benchmark for the complex Langevin procedure. In addition, only when we simulate close enough to the real-time axis, do we have access to these quantities in an undistorted fashion and in turn compute the spectral function of the system.

\subsection{Non-equilibrium dynamics}

Having confirmed the efficacy of the implicit solver for the simulation of the early real-time dynamics of the anharmonic oscillator in thermal equilibrium the next step is to move to an out-of-equilibrium setting. 

We follow ref.\cite{BergesSexty2007} and choose a Gaussian initial density matrix. Its form allows us to incorporate the information about initial conditions into a modification of the action of the system on the first and last point on the Schwinger-Keldysh contour. Note that here the contour consists only of a forward and backward real-time branch, which are not connected via periodic boundary conditions. The most general form of the Gaussian density matrix \cite{Berges2004} leads to the following expression for the system action
\begin{equation}
\begin{aligned}
    S_{\rm G}[\phi_+,\phi_-] =& S[\phi_+] - S[\phi_-] - iS_0(\phi_+[t=0],\phi_-[t=0]) \quad \textrm{with} \\
    S_0[\phi_+,\phi_-] =& i \dot \phi_0 (\phi_+ - \phi_-) - \frac{\sigma^2 + 1}{8\xi^2} \left( \left(\phi_+ - \phi_0\right)^2 + \left(\phi_- - \phi_0\right)^2 \right) \\
    &+ \frac{i\eta}{2\xi}\left( \left(\phi_+ - \phi_0\right)^2 - \left(\phi_- - \phi_0\right)^2 \right) 
    + \frac{\sigma^2 - 1}{4\xi^2}\left(\phi_+ - \phi_0\right)\left(\phi_- - \phi_0\right).
\end{aligned}
\end{equation}
The five independent parameters, which specify the Gaussian initial state, represent the initial values of the field expectation value, the two-point correlation function and their derivatives
\begin{equation}
\begin{aligned}
    \phi_0 =& \langle \phi(t=0) \rangle, \quad \dot\phi_0 = \langle \dot\phi(t=0) \rangle, \\
    \zeta^2 =& \langle \phi(t=0)\phi(t=0) \rangle_c, \\
    \eta\zeta =& \frac12\langle \dot\phi(t=0)\phi(t=0) + \phi(t=0)\dot\phi(t=0) \rangle_c, \\
    \eta^2 + \frac{\sigma^2}{4\zeta^2} =& \langle \dot\phi(t=0)\dot\phi(t=0) \rangle_c. \\
\end{aligned}
\end{equation}
The subscript $c$ refers to the connected correlator, in which the expectation value of the field and its derivatives are subtracted.

The drift term of the discretized complex Langevin dynamics (\cref{eq:discretized_action_derivaitve}) is affected by the Gaussian initial density matrix only at the boundaries of the contour. Consistent with the trapezoidal rule underlying the discretization of the action integral, we choose forward derivatives at the starting point and backward derivatives when considering the endpoint of the contour. No changes are needed at intermediate contour steps. Similar to \cite{BergesSexty2007} we set $\eta=0$ and $\dot \phi_0 = 0$, which leads to the following two explicit terms to implement at the boundary
\begin{equation}
\begin{aligned}
\frac{\delta S_{\rm G}}{\delta \phi_0} =& \frac{1}{|a_0|} \left\{ - \frac{\phi_1 - \phi_0}{a_0} - \frac12 a_0 \frac{\partial V(\phi_0)}{\partial \phi_0} \right. \\  &+\left. \frac12 i \left[ \frac{\sigma^2+1}{4\xi^2} (\phi_0 - \bar\phi)   - \frac{\sigma^2-1}{4\xi^2} (\phi_{N_{\cal C}}-\bar\phi) \right] \right\}, \\
\frac{\delta S_{\rm G}}{\delta \phi_{N_{\cal C}}} =&  \frac{1}{|a_{{N_{\cal C}}-1}|} \left\{ \frac{\phi_{N_{\cal C}} - \phi_{{N_{\cal C}}-1}}{a_{{N_{\cal C}}-1}} - \frac12 \frac{\partial V(\phi_{N_{\cal C}})}{\partial \phi_{N_{\cal C}}} \right. \\ & \left. + \frac12 i \left[ \frac{\sigma^2+1}{4\xi^2} (\phi_{N_{\cal C}} - \bar\phi)   - \frac{\sigma^2-1}{4\xi^2}(\phi_0-\bar\phi) \right] \right\}.
\end{aligned}
\end{equation}
The Langevin equation remains in its standard form for the field degrees on the forward and backward branch respectively
\begin{equation}
    \partial_{\lt} \phi_\pm(x) = i\frac{\delta S_{\rm G}[\phi_+,\phi_-]}{\delta \phi_\pm(x)} + \eta(x,t),
\end{equation}
with no changes to the noise term.

As in previous studies in the literature, we deploy here $m=1$ and a relatively small coupling of $\lambda=1$. This choice leaves us safely in the regime where CLE converges to the right solution. The field starts out at a finite expectation value $\phi_0 = \langle \phi(t=0) \rangle = 1$ at rest $\dot \phi_0 = 0$. The spread in the values of the initial field, encoded in the correlation function is set symmetrically $\sigma=1$ to a value of $\zeta=1$. Mixing terms between field and derivatives vanish via $\eta=0$. We distribute 32 points along each of the two real-time branches to cover the maximum time extent of $x_0^{\rm max}=0.5$.

Similar to the thermal case we deploy the EM solver with $\theta=0.6$ implicitness parameter using the Julia adaptive step size prescription with a maximum Langevin step size of $\Delta \lt=0.005$. Statistics are collected on 500 different trajectories of length $\lt=100$, reading out observables on intervals $\delta \lt=0.1$. Comparisons to matrix mechanics are also available in this scenario, however, the energy eigenfunctions of the harmonic oscillator are not well suited for truncating this particular Hilbert space. Instead, we discretize the Hamiltonian in the coordinate basis using 1024 points in the distance range $\langle x\rangle \in [-10,10]$, the result of which will be shown as solid lines in the subsequent plots.

\begin{figure}
    \includegraphics[scale=0.42]{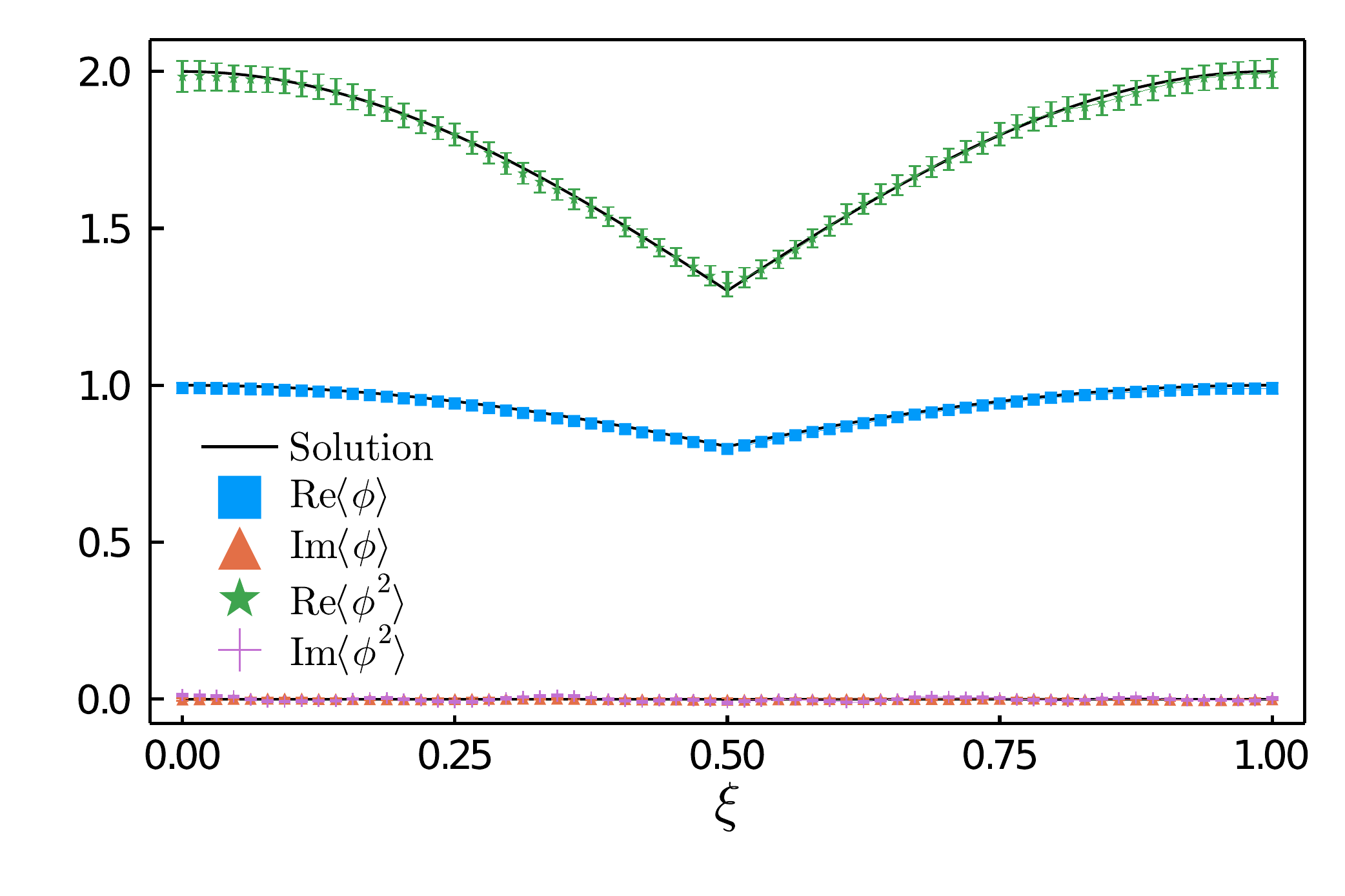}
    \includegraphics[scale=0.42]{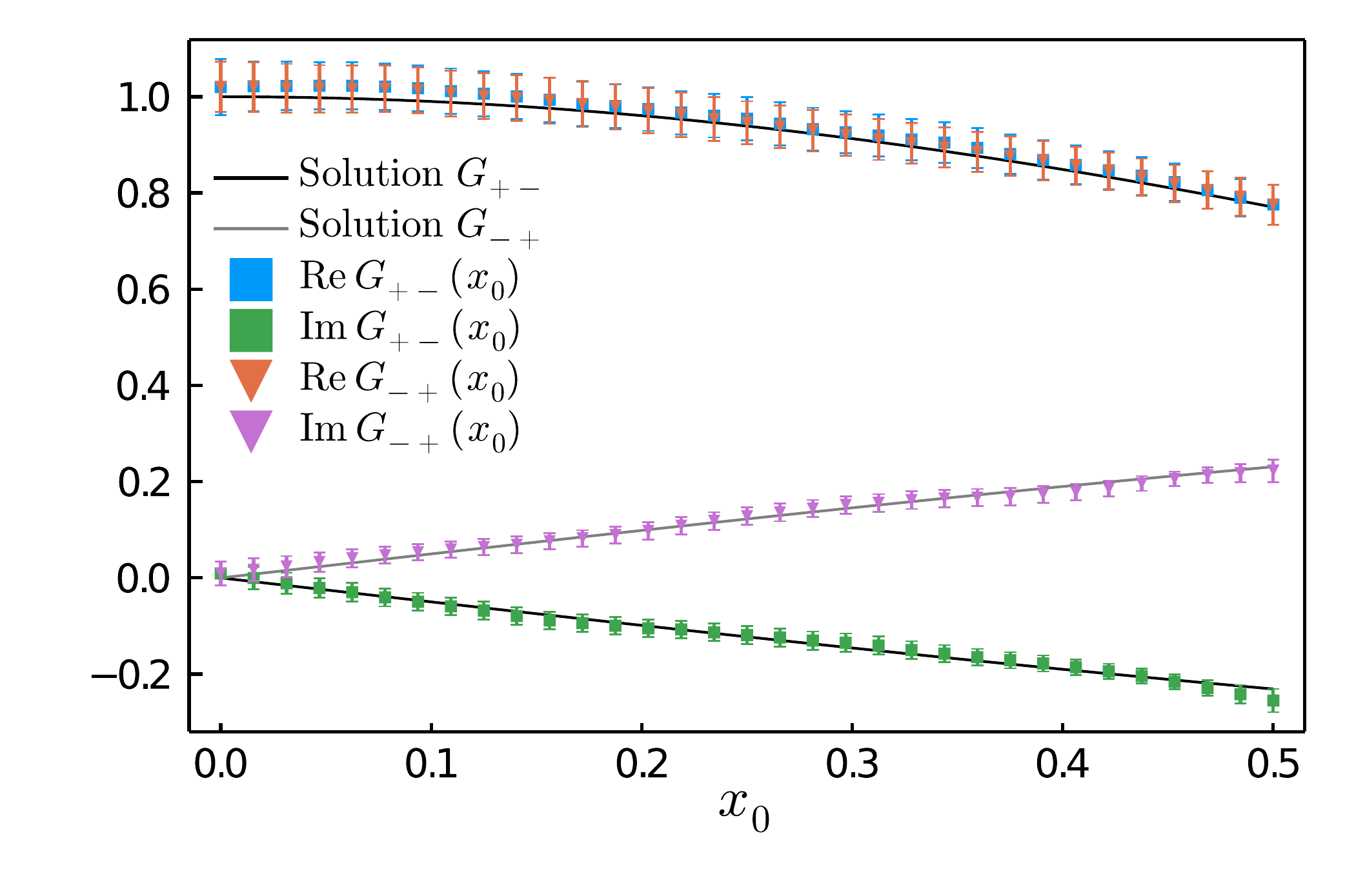}
    \caption{(top) The field expectation value $\langle \phi\rangle$ (box and triangle), as well as the equal-time correlation function $\langle \phi^2\rangle$ (filled star and cross) evaluated out-of equilibrium for a Gaussian density matrix, along the whole simulation contour, parametrized by $\xi$. The continuous Langevin time solution from matrix mechanics is given as solid line. (bottom) The phenomenologically relevant non-equilibrium forward $G_{+-}=G^>$ and backward $G_{-+}=G^<$ correlation functions.}
    \label{fig:nonEquil_corr0t}
\end{figure}

Our out-of-equilibrium simulation results are collected in \cref{fig:nonEquil_corr0t}. Now with time translational invariance gone, we can follow the non-trivial behavior of the field expectation value $\langle \phi\rangle$, plotted as solid squares and triangles in the upper panel along the contour, parameterized by $\xi$. The initial conditions of $\phi_0=1$ as well as unit variance manifest themselves in the value $\langle \phi^2\rangle (\xi=0)=2$. Our results agree within statistical uncertainties with the analytic solution on the real-time axis.

The most interesting unequal-time correlation functions in the out-of-equilibrium scenario are the forward and backward quantities $G_{+-}=G^>$ and $G_{-+}=G^<$, which provide access to the spectral function of the system. Plotted in the lower panel of \cref{fig:nonEquil_corr0t}, we find that here the statistical error remains larger than in the thermal case at the same collected statistics, indicating the presence of larger excursions in the Langevin dynamics as in the more strongly coupled thermal case. Compared to the continuous Langevin time solution from matrix mechanics, the numerical solution again shows excellent agreement. 

\section{Summary and Outlook}
\label{sec:sumout}

In this study, we have explored and showcased the potential of implicit solvers in real-time complex Langevin simulations. With the intention to disentangle the issue of numerical artifacts, such as runaway trajectories, from foundational issues, such as the convergence to wrong results, our focus in this paper remained restricted solely to early real-times.

Two central benefits of the implicit solvers were laid out in detail. On the one hand, the implicit solvers can be shown to be unconditionally asymptotically stable, preventing the occurrence of runaway trajectories, as long as the underlying complex Langevin dynamics remain finite. Using the Langevin dynamics of the free theory as a simple but relevant example, we showed in \cref{sec:langevin_equation} that the difference between implicit and explicit methods lies in the accumulation of errors that either undershoot or overshoot the true trajectory. While the undershoot in the implicit case also leads to a reduction in accuracy of the solution, it manages to prevent the occurrence of runaways.

In \cref{sec:towards} we carried out a comparison of the update prescription for the explicit and implicit EM scheme, which revealed that the effect of the latter can be captured in one additional term in an effective action. That term takes the form of a regulator $+iR$ and depends on the implicitness parameter $\theta$, as well as Langevin step size $\Delta \lt$. Since $R>0$, it indeed dampens the oscillations in the underlying path integral. We conclude that this additional term provides an intrinsic regularization of the path integral unavailable to the explicit solvers.

Subsequently, we analyzed the finite Langevin time discretization artifacts in terms of an effective action in the Fokker-Planck equation for the case of a purely real path integral. We then heuristically generalized the result to the complex case and provided numerical support that our educated guess indeed captures the numerical artifacts introduced due to finite Langevin time steps in the free theory and even the strongly coupled interacting case. This correction formula allows us to exploit the inherent regularization properties of the implicit solvers in practice, as we may now simulate the system at a small but finite Langevin step size $\Delta \lt$ in a well-defined manner and correct for the effect of the regulator a posteriori.

The first three sections have provided us with insight into the regularization properties of different numerical schemes, insight into the effects of finite real-time discretization and we have derived the form of finite Langevin time steps artifacts that allow us to compensate for the effect of the regulator. We thus proceeded in \cref{sec:numsim} to carry out benchmark numerical simulations of the anharmonic oscillator in $(0+1)d$ on the canonical Schwinger-Keldysh contour without tilt and maximum real-time extent of $x_0^{\rm max}=0.5$. Both in the thermal case and in a scenario with Gaussian non-thermal initial conditions, we find excellent agreement between the complex Langevin simulation and the analytic solution from matrix mechanics. The fact that the implicit solver gives access to the backward path on the real-time axis allows us for the first time to compute the actual forward and backward correlators $G_{+-}=G^>$ and $G_{-+}=G^<$, whose difference encodes the phenomenologically relevant spectral function of the system.

We believe that the availability of implicit solvers and an improved understanding of discretization artifacts will help to improve the reliability of the complex Langevin approach and provides new momentum to attack the pressing open challenges associated with it. The stability and regularization properties of the implicit schemes offer benefits in other applications of complex Langevin beyond real-time simulations, such as the treatment of strongly interacting systems at finite chemical potential (for a recent review on CL and the QCD phase diagram see e.g. \cite{Attanasio:2020spv}).

Many different paths forward exist. One aspect we are following up on is the role of regularization in the path integral for the success of complex Langevin convergence. When we introduce a tilt in the Schwinger-Keldysh contour it led us in \cref{eq:tiltreg} to a regulator term that incorporates all terms of the action. We may instead ask how the system reacts to introducing a regulator on individual terms in the action
\begin{eqnarray}
    S &=& i\sum _j [\frac{(\phi _j-\phi_{j-1})^2}{a_j} - a_j \frac{\sigma}{2}\phi _j ^2-a_j \frac{\lambda}{24} \phi _j ^4],
\end{eqnarray}
by modifying in either the kinetic, mass or self-interaction term the lattice spacing from $a_j\to a_j - i\kappa$ with $\kappa>0$. In the following we will thus work with the contour b) of \cref{fig:contours}, where the forward branch is located on the real-time axis and only the backward branch tilts downwards to intersect with the imaginary time axis at $\beta$.

\begin{figure}
        \includegraphics[scale=0.4]{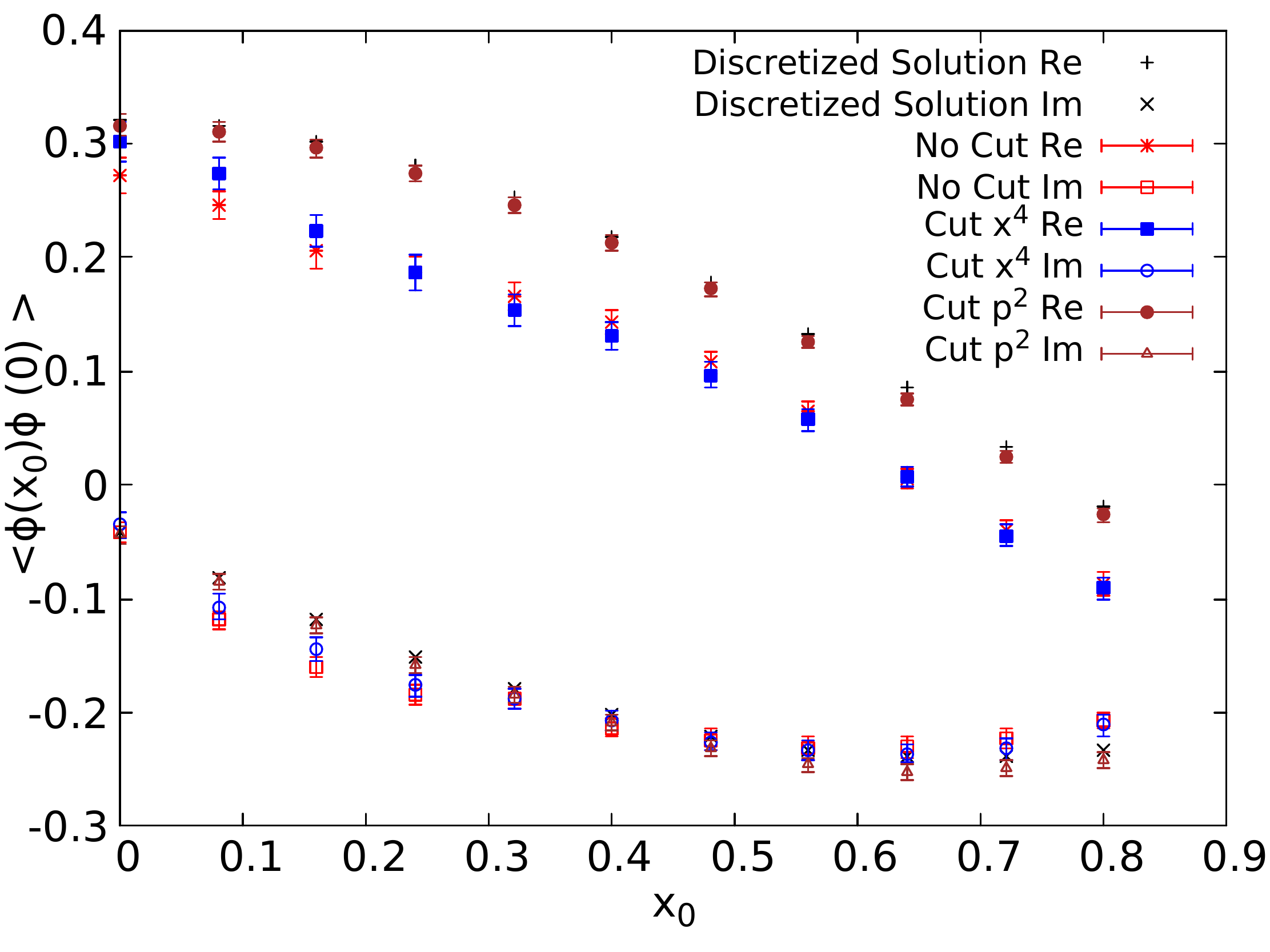}
    \caption{ Comparison of the unequal-time correlation function $G_{++}$ in thermal equilibrium from matrix mechanics (black crosses) with complex Langevin simulations carried out on a Schwinger-Keldysh contour of intermediate real-time extent $x_0^{\rm max}=0.8$. Here the forward branch of the contour resides on the real-time axis and the backward contour tilts down to intersect with the imaginary axis at $\beta$ (c.f. b) in \cref{fig:contours}). The direct simulation based on the explicit EM scheme converges to an incorrect result given by the red data points. No improvement is observed for regularizing the $\phi^4$ term (blue). The correct solution is recovered when regularizing the momentum term (brown).} \label{fig:indivregconv}
\end{figure}

We have seen that for $x_0^{\rm max}=0.5$ the complex Langevin approach in the strongly coupled thermal scenario with $\lambda=24$ converges to the correct solution given by matrix mechanics. Extending the contour to later real-times, we encounter significant deviations already at $x_0^{\rm max}=0.8$. A prominent characteristic of the incorrect solution is an artificial downward shift in the real-part of the unequal-time correlation function, as shown by the red data points in \cref{fig:indivregconv}. In addition, the curvature of the imaginary part of the correlator, given as open squares also deviates from the true solution beyond statistical uncertainty.

In the regime $0.75 < x_0^{\rm max} \lesssim 1$ we observe that this incorrect convergence can be overcome by a choice of regularization on the forward branch. Interestingly, when introducing an imaginary part in the $a_j$'s associated with the interaction term (blue filled square and open circle) the simulation outcome remains unchanged. On the other hand, modifying the kinetic term with a small imaginary part $a_j-i\times 10^{-3}$ leads to a significant improvement as indicated by the brown-filled circle and open triangles, which agree with the analytic solution from matrix mechanics. On the other hand, such a regularization based strategy fails to achieve its purpose, once the real-time extent of the Schwinger-Keldysh contour goes beyond unity in units of the mass. Our goal in future work is to gain a systematic understanding of how the regularization achieves to recover the correct results, possibly by studying the associated Fokker-Planck equation in low-dimensional models. 

Furthermore, the availability of implicit and in particular higher-order solvers benefits the systematic exploration of kernels for the Langevin dynamics (for a modern perspective on CL kernels see e.g. \cite{Aarts:2012ft}). In the real-valued case, kernels can be used to improve the convergence properties of the stochastic quantization procedure. In complex Langevin, they have been studied with mixed success as means to remedy the convergence to wrong solutions. Robust numerical SDE solvers (c.f. the Runge-Kutta Milstein scheme of \cref{eq:RKMilScheme}), which can accommodate non-trivial kernels with Langevin-time and field dependencies will allow us to explore a much broader class of kernels than before in future studies.

\section*{Acknowledgements}
The team of authors gladly acknowledges support by the Research Council of Norway under the FRIPRO Young Research Talent grant 286883. The numerical simulations have been partially carried out on computing resources provided by  
UNINETT Sigma2 - the National Infrastructure for High Performance Computing and Data Storage in Norway under project NN9578K-QCDrtX "Real-time dynamics of nuclear matter under extreme conditions"  


\FloatBarrier

\begin{backmatter}

\section*{Competing interests}
  The authors declare that they have no competing interests.

\section*{Author's contributions}
    \begin{itemize}
        \item D. Alvestad: code development (implicit), data analysis, analytic computations (error estimated), writing
        \item R. Larsen: code development (explicit), data analysis, analytic computations (regularization), writing
        \item A. Rothkopf: project seeding, funding acquisition, supervision, writing
    \end{itemize}


\bibliographystyle{stavanger-mathphys}


\bibliography{references}


\end{backmatter}


\end{document}